\pdfoutput=1
\documentclass[preprint,aps,showpacs,preprintnumbers,amsmath,amssymb]{revtex4}
\usepackage{graphicx}
\usepackage{dcolumn}
\usepackage{bm}
\usepackage{color}
\begin{document}
\title{Spectral and transport properties of the \\two-dimensional Lieb lattice}
\author{M. Ni\c t\u a$^{1}$, B. Ostahie$^{1,2}$ and A. Aldea$^{1,3}$}
\affiliation{$^1$National Institute of Materials Physics, POB MG-7, 
77125 Bucharest-Magurele, Romania. \\
$^2$Department of Physics, University of Bucharest\\
$^3$Institute of Theoretical Physics, Cologne University, 50937
Cologne, Germany.}
\date{\today}
\begin{abstract}
The specific topology of the line centered square lattice (known also as the 
Lieb lattice) induces remarkable spectral properties as the macroscopically 
degenerated zero energy flat band, the Dirac cone in the low energy spectrum,
 and the peculiar Hofstadter-type spectrum in  magnetic field.
We study here the properties of the finite Lieb lattice with periodic
and vanishing boundary conditions. We find out the  behavior of the 
flat band induced by  disorder and external magnetic and electric fields.
We show that in the confined Lieb plaquette threaded by a  perpendicular 
magnetic flux there are edge states with nontrivial behavior.
The specific class of twisted edge states, 
which have  alternating chirality, are sensitive to disorder and do
not support IQHE, but contribute to the longitudinal resistance.
The  symmetry of the transmittance matrix in the energy range
where these states are located is revealed.
The diamagnetic moments of the bulk and edge states in the Dirac-Landau domain, and also of the flat states  in crossed magnetic and electric fields are shown.  
\end{abstract}

\maketitle
\section{ Introduction }
The interest in the line centered square lattice, 
known as the 2D Lieb lattice, 
comes from the specific properties induced by its topology.
The lattice is characterized by a unit cell containing three atoms, and
a one-particle energy spectrum showing  a three band structure
with electron-hole symmetry, one of the branches being
flat and macroscopically degenerate. For the infinite lattice,
the three energy bands touch each other at the  middle of the spectrum
(taken  as the  zero energy ), and the low energy spectrum exhibits
a  Dirac cone located at the point $\Gamma=(\pi,\pi)$  in the 
Brillouin zone.
Except for the presence of the flat band,  the Lieb lattice shows
similarities with the honeycomb lattice in what concerns both  spectral
and transport properties. For instance, besides the presence of the
Dirac cone, the energy spectrum in  the presence of the magnetic field 
shows also a double Hofstadter picture, with the typical $\sqrt{B}$ 
dependence of the relativistic bands on the magnetic field $B$ \cite{Rammal}.
The Hall resistance of the two systems in the quantum regime behaves  alike,
but  the step between consecutive plateaus equals $h/e^2$ in the Lieb case
(instead of $h/2e^2$ for graphene) because of the presence of a single Dirac cone per BZ.
An all-angle Klein transmission is proved by the relativistic electrons 
in the Lieb lattice \cite{Aoki, Goldman, Shen}.

There are more lattices that support flat bands, however it is specific to
the Lieb lattice that the band is robust against the magnetic field, 
while  other lattices develop dispersion at any $B\ne 0$.
The intrinsic spin-orbit coupling  does not  affect the flat band either, 
but opens a gap at the touching point $\Gamma$, the Lieb system becoming 
in this  way a quantum spin Hall phase \cite{Goldman, Weeks}. 
Topological phase transitions driven by different parameters are 
studied in \cite{Smith, Tsai}.
The zero-energy flat bands  became a topic of intense study also
for other reasons: they  may allow for the non-Abelian FQHE \cite{Wang,
DasSarma, Bergholtz} or for ferromagnetic order and surface superconductivity 
\cite{Zhang,Zhao,Volovik}.  

In this paper we address the properties of the {\it finite (mesoscopic)} Lieb
lattice with emphasis on some  features of the flat band and
of the edge states which are specific to this  lattice.
We adopt the spinless tight-binding approach, and the spectral properties are
examined under both periodic and vanishing boundary conditions
applied to the system described in Fig.1.
In section II we find  that the zero energy flat band exists independently 
of the boundary conditions. It turns out, however, that in the  periodic case
the band is built up only from B- and C- orbitals, while
in the other case the A-type orbitals are also involved (see Fig.1).
We prove this analytically   by calculating  the eigenfunction
in both situations. In this way  we also find out that 
for confined  systems (i.e., with vanishing boundary conditions)
the  degeneracy of the flat band  equals $N_{cell}+1$ ($N_{cell}$ is 
the number of cells of the mesoscopic plaquette). We find in section IIIA
that for a confined plaquette two levels separate from the bunch when 
a perpendicular magnetic field is applied, such that the degeneracy is 
reduced by 2. This is proved  in a perturbative manner for the general case, 
however it can be observed more easily by the use of the toy model consisting 
of two cells only.
		                                                    
Next, we study  how the flat band degeneracy is lifted by disorder
and by an external electric field applied in-plane.
An exotic result is that the extended  states of the disordered flat band 
in the presence of a magnetic field behave according to the orthogonal 
Wigner-Dyson distribution although the unitary distribution is expected.
When an electric field is applied, the flat band splits in a Stark-Wannier
ladder  whose structure is analyzed by calculating the diamagnetic moments 
of the states in crossed electric and magnetic fields.

In section IIIB we study the edge states which fill the gaps of the
double Hofstadter butterfly when the magnetic field is applied on
the confined Lieb plaquette. We identify three types of such states. 
The conventional edge states located between the Bloch-Landau bands 
and also between Dirac-Landau bands (i.e., the relativistic range
of the spectrum) differ, as expected, 
in their chirality.  Additionally, we detect {\it twisted edge states} 
situated in the magnetic gap which protect the zero-energy band, 
coming in bunches and characterized  by an oscillating chirality 
as function of the magnetic field.
The twisted edge states show remarkable  properties: surprisingly,
they are not robust to disorder, as the other types of edge states are, 
and does not carry transverse current (i.e., the QHE vanishes in the energy 
range covered by these states).
The last property comes from a specific symmetry of the transmittance
matrix which is discussed in section IV.

Finally, one has to note that  the line centered square lattices are  found 
in nature as  $Cu-O_2$ \cite{Lieb} planes in cuprate superconductors, 
and can be engineered as an optical lattice \cite{Goldman,Manninen}.
\section{The tight-binding model for the Lieb lattice :
periodic versus vanishing  boundary conditions}

Our aim is to point out  specific aspects of the confined 
Lieb plaquette from the point of view of spectral and transport properties.
In order to allow for a comparison we shortly  describe also  the  
case of the infinite system, with and without magnetic field, although the
eigenvalue problem is already known from the literature.
We remind  that the continuous model for the infinite Lieb system
in perpendicular magnetic field \cite{Goldman}, shows the $\sqrt {B}$ 
dependence on the magnetic field of the eigenenergies in the relativistic 
range. 
The information obtained in the long-wave approximation of the Schr\"{o}dinger
equation, concerning  the dependence on $B$ of the Bloch-Landau or
 Dirac-Landau bands are recaptured  in the spectrum of the discrete 
 tight-binding model (Fig.5a) together with effects coming from the periodic 
lattice and finite edges.

In this section, starting from the tight-binding Hamiltonian,  we built up 
the eigenfunction of the periodic and finite Lieb plaquette, and prove the
degeneracy and structure of the zero-energy flat band. The crossover from
the simple Hofstadter spectrum of the simple square lattice to the
Lieb spectrum characterized by a double butterfly, magnetic gap and a flat
band is shown in Fig.3.

\begin{figure}[htb]
        \includegraphics[angle=-0,width=0.50\textwidth]{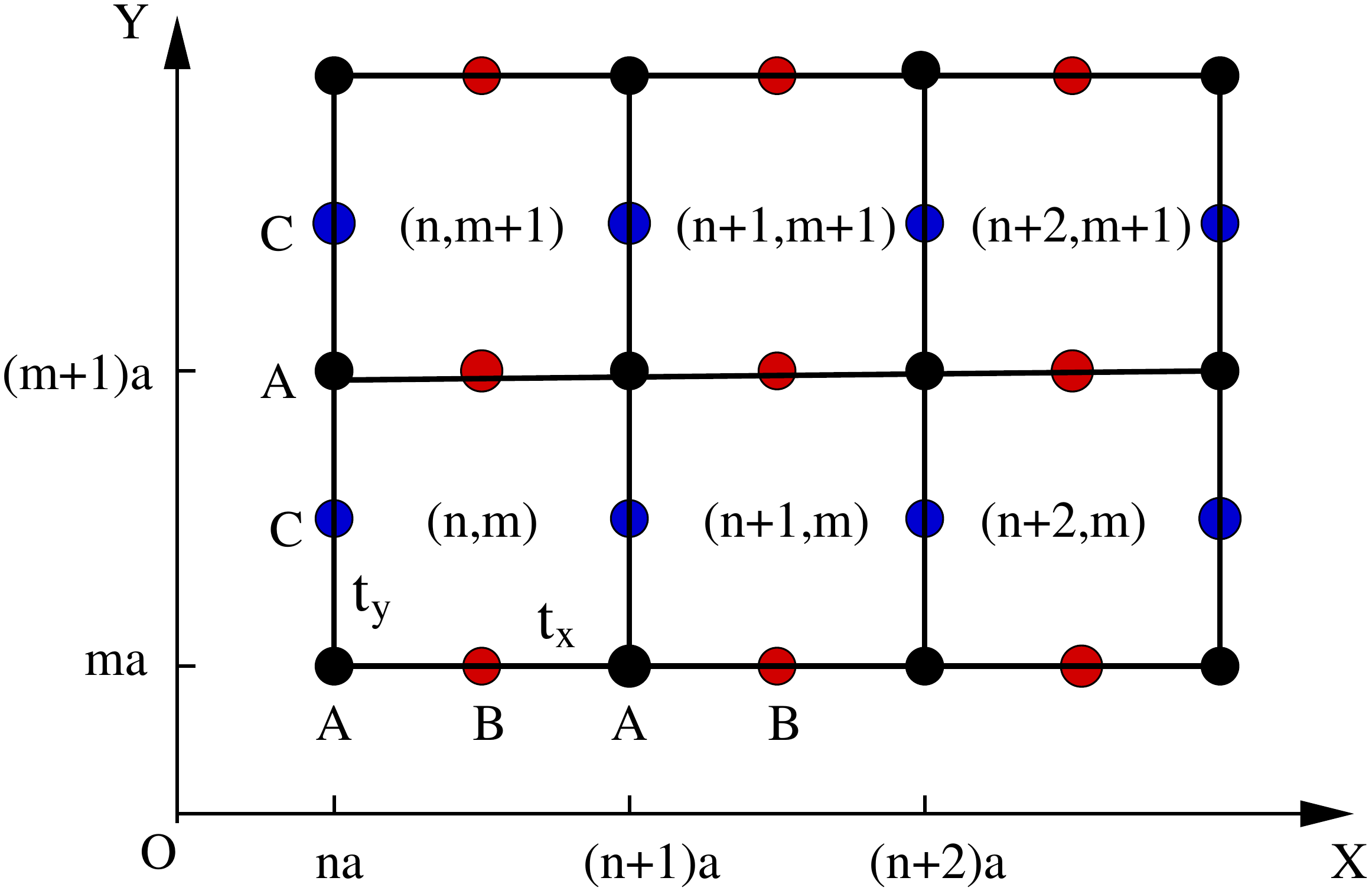}
\caption{(Color online) The Lieb lattice: the unit cell contains three atoms $A,B$ and $C$; indices (n,m)  identify the cell; $t_x,t_y$ are the hopping
integrals along the directions $Ox$ and $Oy$, respectively; 
$a$ is the lattice constant.}
\end{figure}
The Lieb lattice is a 2D square lattice with centered lines as shown in Fig.1.
It is characterized by three atoms ($A,B,C$) per unit cell, 
the connectivity of the atom $A$ being equal to four, while  the connectivity 
of atoms $B$ and $C$  equals two.

Introducing creation $\{a^\dagger_{nm},b^\dagger_{nm},c^\dagger_{nm}\}$
and annihilation  $\{a_{nm},b_{nm},c_{nm}\}$ operators of the 
localized states $|A_{nm}>, |B_{nm}>, |C_{nm}>$
~(where $(nm)$ stands for the cell index and the letters $A,B,C$ identify the
type of atom),~
the spinless tight-binding Hamiltonian of the Lieb lattice in perpendicular 
magnetic field reads:
\begin{eqnarray}
  \nonumber &H=\sum_{nm} E_a a^{\dagger}_{nm}a_{nm}+
  E_b b^{\dagger}_{nm}b_{nm}+E_c c^{\dagger}_{nm}c_{nm}\\
  & +t_x e^{-i\pi  m\phi} a^{\dagger}_{nm}b_{nm}+
  t_x e^{i \pi m\phi} a^{\dagger}_{nm}b_{n-1,m}
  +t_y a^{\dagger}_{nm}c_{nm}+ t_y a^{\dagger}_{nm}c_{n,m-1} \nonumber \\
  & +t_x e^{-i\pi m\phi}b^{\dagger}_{nm}a_{n+1,m}+
  t_x e^{i\pi m\phi}b^{\dagger}_{nm}a_{nm}
   +t_y c^{\dagger}_{nm}a_{nm}+t_yc^{\dagger}_{nm}a_{nm+1} ,
  \end{eqnarray}
where $\phi$ is the flux through the unit cell of the Lieb lattice 
measured in quantum flux units; we mention that 
the vector potential has been chosen as $\vec{A}=(-By,0,0)$.

The presence of a spectral flat band can be noticed already in the simplest 
case  of the { \it periodic} boundary conditions and vanishing magnetic flux.
Assuming that the lattice is composed of $N_{cell}^x=N$ cells along $Ox$ and  
$N_{cell}^y=M$ cells along  $Oy$, the Fourier transform 
$c_{\vec{k}}= c_{k_x,k_y}= 
\frac{1}{\sqrt{NM}}\sum_{n,m} c_{nm}e^{i(k_xn+k_ym)}$
~(and similarly for all the other operators)  
yields the $k$-representation of the Hamiltonian 
described by a  $3\times 3$ matrix:
\begin{equation}
          H=\sum_{\vec{k}}
    \left(\begin{array}{ccc}
      a^{\dagger}_{\vec{k}}~ & b^{\dagger}_{\vec{k}} ~& c^{\dagger}_{\vec{k}}
\end{array}\right)
        \left(\begin{array}{ccc}
                E_a & \Delta^*(k_x) & \Lambda^* (k_y) \\
                \Delta(k_x) & E_b & 0  \\
                \Lambda(k_y) & 0 & E_c
        \end{array}\right)
    \left(\begin{array}{c}
            a_{\vec{k}} \\
            b_{\vec{k}}  \\
            c_{\vec{k}} 
\end{array}\right),
\end{equation}
where $k_x=2\pi p/N~ (p=1,..,N)$, $k_y=2\pi q/M~ (q=1,..,M)$, and the
 notations $\Delta(k_x)=t_x(1+e^{ik_x})$ , 
$\Lambda(k_y)=t_y(1+e^{ik_y})$ has been used.
With the choice  $E_a=E_b=E_c=0$, one obtains the following eigenvalues:
\begin{eqnarray}
\nonumber        \Omega_{\pm}(\vec{k})=&&\pm\sqrt{|\Delta|^2+|\Lambda|^2}\
	=\pm 2\sqrt{t_x^2 cos^2(k_x/2)+t_y^2cos^2(k_y/2)},~ ~\\
	\Omega_0(\vec{k})=&&0 ,
\end{eqnarray}
where $\Omega_{\pm}$ are the energies of the upper and lower band,
respectively, and $\Omega_0$ is   the  nondispersive (flat) band of the
Lieb lattice. 
The most interesting point in the BZ  is the  point
 $\Gamma=(\pi,\pi)$, 
where  in the case of the infinite lattice the three branches are touching 
each other. The expansion of the functions $\Delta(k_x)$ and $\Lambda(k_y)$ 
about this point  gives rise to a Dirac cone (massless) spectrum:
\begin{equation}
\Omega_{\pm}=\pm\sqrt{t_x^2 k_x^2 + t_y^2 k_y^2}.
\end{equation}
On the other hand, the expansion of the same functions
about $R=(0,0)$ shows a parabolic  dependence:
\begin{equation}
\Omega_{\pm}=\pm \big(\frac{k_x^2}{2m_x} + \frac{k_y^2}{2m_y}\big),
\end{equation}
where  $m_x,m_y$ are effective masses along the two directions.

Other relevant points in the BZ are $M=(\pi,0)$ and $(0,\pi)$, which 
prove to be saddle points in the spectrum as it can be noticed also in Fig.2.
Above and below the corresponding energy  $E=\pm 2 t$ 
(where we considered  $t_x=t_y=t$) the effective mass exhibits opposite signs 
inducing the change of sign of the Hall effect which is visible in Fig.15.

For comparison's sake, we remind that the energy spectrum of the honeycomb 
lattice contains two cones per BZ, and that the saddle point occurs at the 
energy $E=\pm t$. The tight-binding spectrum of the graphene extends over the 
interval [-3t,3t], while for the Lieb lattice the interval is [$-2\sqrt{2}t,
2\sqrt{2}t$].
\begin{figure}[htb]
\includegraphics[angle=-00,width=0.35\textwidth]{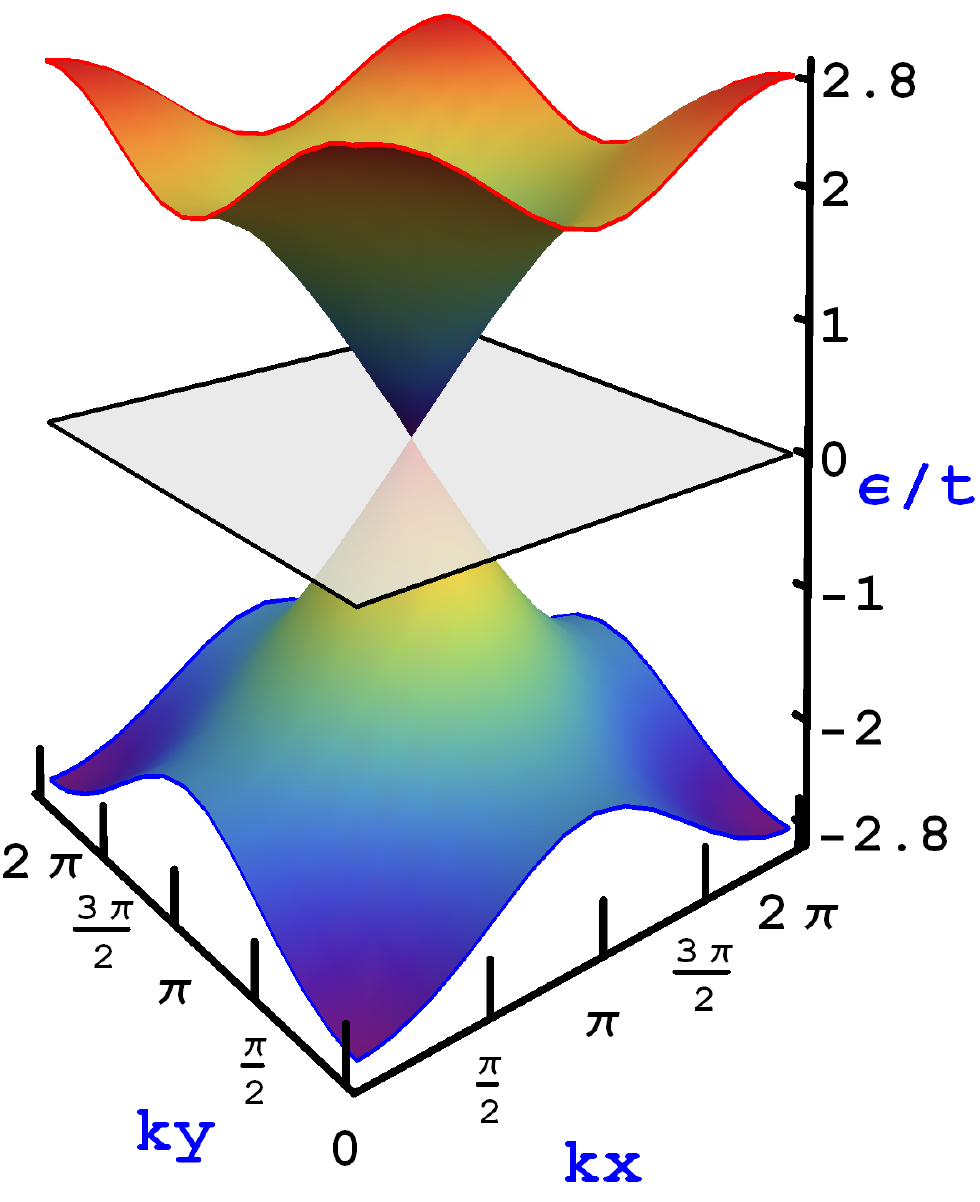}
\hskip0.2cm
\includegraphics[angle=-00,width=0.35\textwidth]{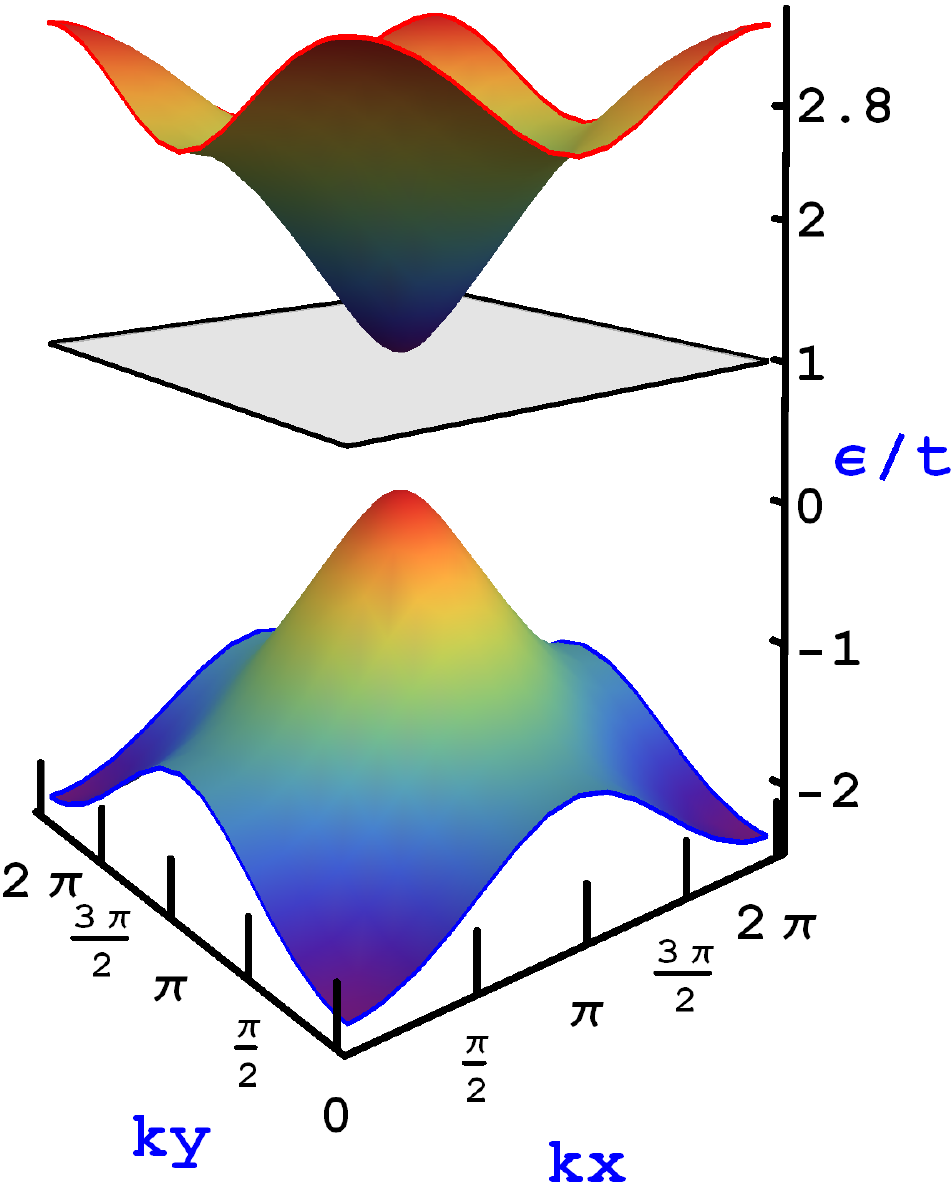}
\caption {(Color online) The energy spectrum 
of the infinite Lieb lattice.
(left) The case $E_a=E_b=E_c=0$ when the three bands 
(two dispersive and one flat) get in contact at 
$\vec{k}=(\pi,\pi)$. At low energy the dispersion is linear giving rise 
to Dirac cones. (right) The staggered case $E_a=0,E_b=E_c=1$ when
the spectrum is gapped and the rounding of cones is obvious. }
\end{figure}

In order to get supplementary information about the origin of the flat band
let us consider the staggered case $E_a=0,E_b=E_c=E_0$.
Then,  a gap is expected in the energy spectrum, and, indeed, the  
eigenvalues are  now \cite{Shen}:
\begin{eqnarray}
	\Omega_{\pm}(\vec{k})=\frac{1}{2}\big[E_0\pm\sqrt{E_0^2+
	4(|\Delta|^2+|\Lambda|^2)}~\big]
	,~ ~\Omega_0(\vec{k})= E_0.
\end{eqnarray}  
The new  spectrum is shown in Fig.2b, where one notices the persistence of
the flat band, which is however shifted to $E=E_0$. Since the energy $E=E_0$
corresponds to the atomic level of the orbitals $B$ and $C$, the result
argues that the flat band states are created  only by this type of orbitals.
The gap induced by the staggered arrangement is accompanied by the rounding 
of the cones, that indicates a non-zero effective mass in the low-energy 
range of the two spectral branches, as it can be observed in Fig.2b. 

In what follows we shall calculate the eigenfunctions 
of the {\it finite} Lieb lattice, imposing first periodic 
conditions, and then the  vanishing boundary conditions proper  to the confined
plaquette.
Let $\Psi_{\vec{k}}$ be the eigenfunctions of the Lieb lattice with periodic
boundary conditions built up as the linear combination:
\begin{equation}
\Psi_{\vec{k}}= \alpha_{\vec{k}}a^{\dagger}_{\vec{k}}|0>+
\beta_{\vec{k}}b^{\dagger}_{\vec{k}}|0>+
\gamma_{\vec{k}}c^{\dagger}_{\vec{k}}|0> ,
\end{equation}
where the coefficients $\alpha_{\vec{k}},\beta_{\vec{k}},\gamma_{\vec{k}}$ 
satisfy the equations:
\begin{eqnarray} 
\nonumber E_a\alpha_{\vec{k}}+&\Delta^*(k_x)\beta_{\vec{k}}&+
\Lambda^*(k_y)\gamma_{\vec{k}}=E \alpha_{\vec{k}}\\
\nonumber &\Delta(k_x) \alpha_{\vec{k}}&+E_b\beta_{\vec{k}}=
E \beta_{\vec{k}}\\
&\Lambda(k_x) \alpha_{\vec{k}}&+E_c\gamma_{\vec{k}}=E \gamma_{\vec{k}}.
\end{eqnarray}
Then, the functions corresponding to the eigenvalues $\Omega_0$ 
 and $\Omega_{\pm}$ in Eq.(3) read:
\begin{eqnarray} 
& \Psi^0(\vec{k})&=\frac{1}{\sqrt{|\Delta|^2+|\Lambda|^2}} \big(
\Lambda^*(k_y)b^{\dagger}_{\vec{k}}-\Delta^*(k_x) c^{\dagger}_{\vec{k}}\big)|0>\\
&\Psi^{\pm}(\vec{k})&=\frac{1}{2}\big(\pm a^{\dagger}_{\vec{k}}
+\frac{\Delta(k_x)}{{\sqrt{|\Delta|^2|+|\Lambda|^2}}}~b^{\dagger}_{\vec{k}}+
\frac{\Lambda(k_y)}{{\sqrt{|\Delta|^2 +|\Lambda|^2}}}~
c^{\dagger}_{\vec{k}}\big)|0> .
\end{eqnarray}

In the case of periodic conditions applied to the finite plaquette
there are some subtleties concerning the band degeneracy which become
unimportant in the limit of infinite system. It is obvious 
from Eqs.(9-10) that the three bands come into contact at $\vec{k}=(\pi,\pi)$, 
however this value of $\vec{k}$ is allowed only if both $N$ and $M$ are even. 
In this case the flat band at $E=0$ is ($N_{cell}+2$)- fold degenerate, 
otherwise all the three bands are $N_{cell}$-fold degenerate 
(where the number of cells $N_{cell}=NM$).

The expression of $\Psi^0(\vec{k})$ in Eq.(9) indicates again that the 
flat band of the periodic lattice
is composed only from orbitals of the type $B$ and $C$. 
On the other hand, we shall see below that in the case of 
vanishing boundary conditions the
zero-energy eigenfunction may sit also on the $A-$type sites, and 
that the degeneracy of the flat band becomes $N_{cell}+1$.

The periodic boundary conditions can be used 
in the presence of a uniform perpendicular magnetic field
 for rational values of the magnetic flux $\phi=p/q$
resulting in a spectrum composed of two Hofstadter butterflies similar 
 to  the case of the honeycomb lattice. 
However, in contradistinction to the honeycomb lattice, one notices  the 
presence of a dispersionless band at $E=0$, which is flat with respect  
to the variation of the magnetic flux, and is protected by a gap opened at 
$B\ne0$ \cite{Aoki,Goldman} . 
The spectrum exhibits  Bloch-Landau  bands at the 
extremities  and also relativistic Dirac-Landau bands towards the middle.
The two types of bands are distinguished by  opposite chirality $dE/d\phi$
and by different dependence on the magnetic field.

The periodic boundary conditions discussed above can be properly 
used  for describing infinite lattices, however when interested in 
mesoscopic plaquettes they have to be replaced with vanishing boundary 
conditions. 
We intend to identify the differences introduced by the {\it finite size}, 
which will turn out to be non-trivial in the case of the Lieb lattice.

For the confined Lieb lattice, the eigenfunctions 
can be obtained as  combinations of  functions Eq.(9) or Eq.(10) with 
coefficients that ensure the vanishing of the eigenfunction along the  edges. 
As a technical detail we mention that (along the Ox direction, for instance)
the finite plaquette begins with the atom $A$ in the first cell, and also
ends   with an atom $A$ which belong to 
 the $(N+1)$-th cell. This means that
the  wave function $|\Phi(\vec{k})>$  should vanish  at the site 
$B$ in the $0$-th and $(N+1)$-th  cell, i.e.:
$<\Phi(\vec{k})|b^{\dagger}_{N+1,m}|0>=<\Phi(\vec{k})|b^{\dagger}_{0,m}|0>=0$. 
Similarly,  the vanishing condition along Oy occurs at
the site $C$ in the $0$-th and  $(M+1)$-th cell along this direction, i.e.:
$<\Phi(\vec{k})|c^{\dagger}_{n,M+1}|0>=<\Phi(\vec{k})|c^{\dagger}_{n,0}|0>=0$. 

In the localized representation, which is the proper one in the case
of confined systems, the eigenfunctions $|\Phi^0(\vec{k})>$
corresponding to $E=\Omega_0=0$  look as follows:
\begin{eqnarray}
\nonumber|\Phi^0(\vec{k})>= \sqrt{\frac{2}{N+1}}\sqrt{\frac{2}{M+1}}
\sum_{n=1}^{N+1}\sum_{m=1}^{M+1} 
\big(\frac{2t_y cos\frac{k_y}{2}}{\sqrt{|\Delta|^2+|\Lambda|^2}} 
sin k_x n~ sin k_y(m-\frac{1}{2})~ b^{\dagger}_{nm}|0>\\
-\frac{2 t_x cos\frac{k_x}{2}}{\sqrt{|\Delta|^2+|\Lambda|^2}}
sin k_x(n-\frac{1}{2})~ sin k_ym~ c^{\dagger}_{nm}|0>\big),
\end{eqnarray}
where $k_x,k_y$ are obtained from the condition that the wave function 
vanishes at the boundary, and equal $k_x= p\pi/(N+1)$~ (p=1,..,N+1)
and $k_y= \frac{q\pi}{M+1}$~ (q=1,..,M+1).
Since the situations $p=N+1$ (at any $q$) and $q=M+1$  (at any $p$),
generate $|\Phi^0>=0$, we are left in Eq.(11) with  only $NM $  
non-vanishing degenerate orthogonal  eigenfunction.

The eigenfunctions $|\Phi^{\pm}(\vec{k})>$ corresponding to the other 
 two energy branches can be written similarly as:
\begin{eqnarray}
|\Phi^{\pm}(\vec{k})>= \sqrt{\frac{2}{(N+1)(M+1)}}
\sum_{n=1}^{N+1}\sum_{m=1}^{M+1}\big(\pm sin k_x(n-\frac{1}{2})
sin k_y(m-\frac{1}{2}) ~a^{\dagger}_{nm}|0> \nonumber\\
+\frac{2t_x cos \frac{k_x}{2}}{\sqrt{|\Delta|^2+|\Lambda|^2}}
sin k_x n~ sin k_y(m-\frac{1}{2})~ b^{\dagger}_{nm}|0> \nonumber\\
+\frac{2t_y cos \frac{k_y}{2}}{\sqrt{|\Delta|^2+|\Lambda|^2}}
sin k_x(n-\frac{1}{2}) sin k_ym~ c^{\dagger}_{nm}|0>\big) ,
\end{eqnarray}
where states of the type $A$ are this time also present.
One can readily see that the number of non-vanishing states in each 
spectral branch is $(N+1)(M+1)-1$, since the  point 
$\Gamma=(\pi,\pi)$ has to  be treated separately. This is  because its
corresponding energy vanishes  and the state should be counted in the
flat band. In this case the wave function becomes:
\begin{equation}
|\Phi^0_a>=:|\Phi^{\pm} (\pi,\pi)>=\sqrt{\frac{1}{N+1}}\sqrt{\frac{1}{M+1}}
\sum_{n=1}^{N+1}\sum_{m=1}^{M+1} (-1)^{n+m}~ a^{\dagger}_{nm}|0> .
\end{equation}	 
For the finite Lieb plaquette with vanishing boundary conditions, 
one may conclude that the flat band degeneracy 
equals $NM+1$, while each other branch contains 
$NM+N+M$ states, so that the total number of states equals indeed the
number of sites  $3NM+2(N+M)+1$.

In the presence of the magnetic field,
the vanishing  boundary conditions give rise to edge states which 
fill the gaps of the Hofstadter spectrum corresponding to the periodic system.
Besides the  edge states existing in the energy range of 
the Bloch-Landau levels (which are  the only
met for the  finite plaquette with simple square structure), 
there are  edge states  in  the relativistic 
range which show  opposite  chirality \cite{note3}, but also
non-conventional edge states lying in the central gap  which protects the 
zero energy dispersionless  band. 
This last new class of edge states exhibits oscillating chirality when changing
either the magnetic flux or the Fermi energy. These states will be studied 
in the next chapter.
The fate of the zero-energy states in the presence of confinement will 
be discussed in the next section.

The Lieb lattice can be generated from the simple square lattice  by extracting each the second atom when moving along both $Ox$ and $Oy$ direction.
Formally, this means either to push to infinity the energy $E_d$ of these atoms
or to cut down the hopping integrals $t'$ connecting them to the neighboring 
atoms, and it is instructive to follow the change of the spectrum 
when $E_d/t'\rightarrow \infty$. By driving the system in this way  
from $1$ to $3$ atoms/unit cell, the lattice periodicity is doubled along 
both directions, and the flat band is generated. 
 The middle panel of Fig.3 shows how the butterfly wings
break off during the process giving rise to the relativistic range.
\vskip-5cm
\begin{figure}[htb] 
\hskip-0.2cm
\includegraphics[angle=-00,width=0.4\textwidth]{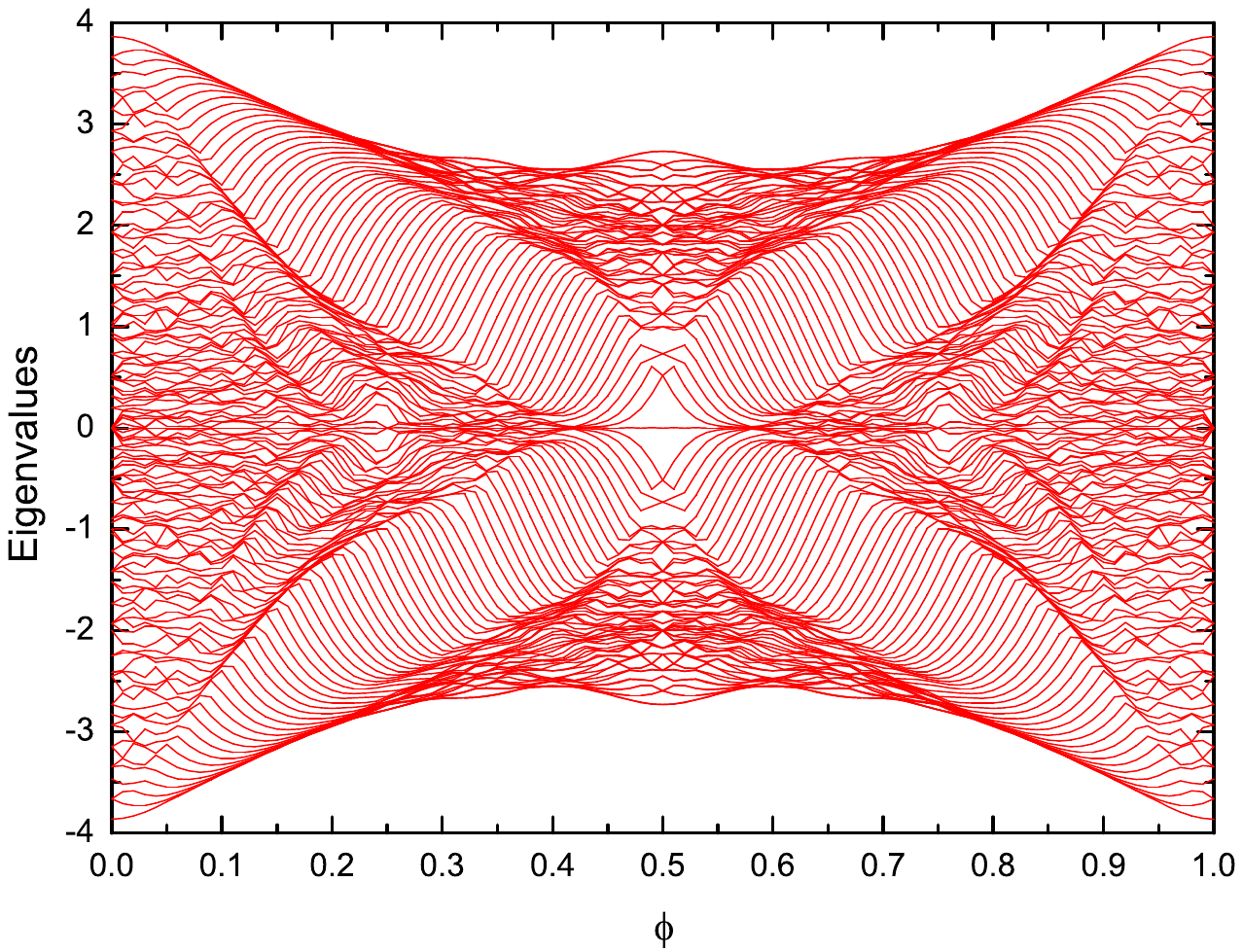}
\hskip-1.8cm
\includegraphics[angle=-00,width=0.4\textwidth]{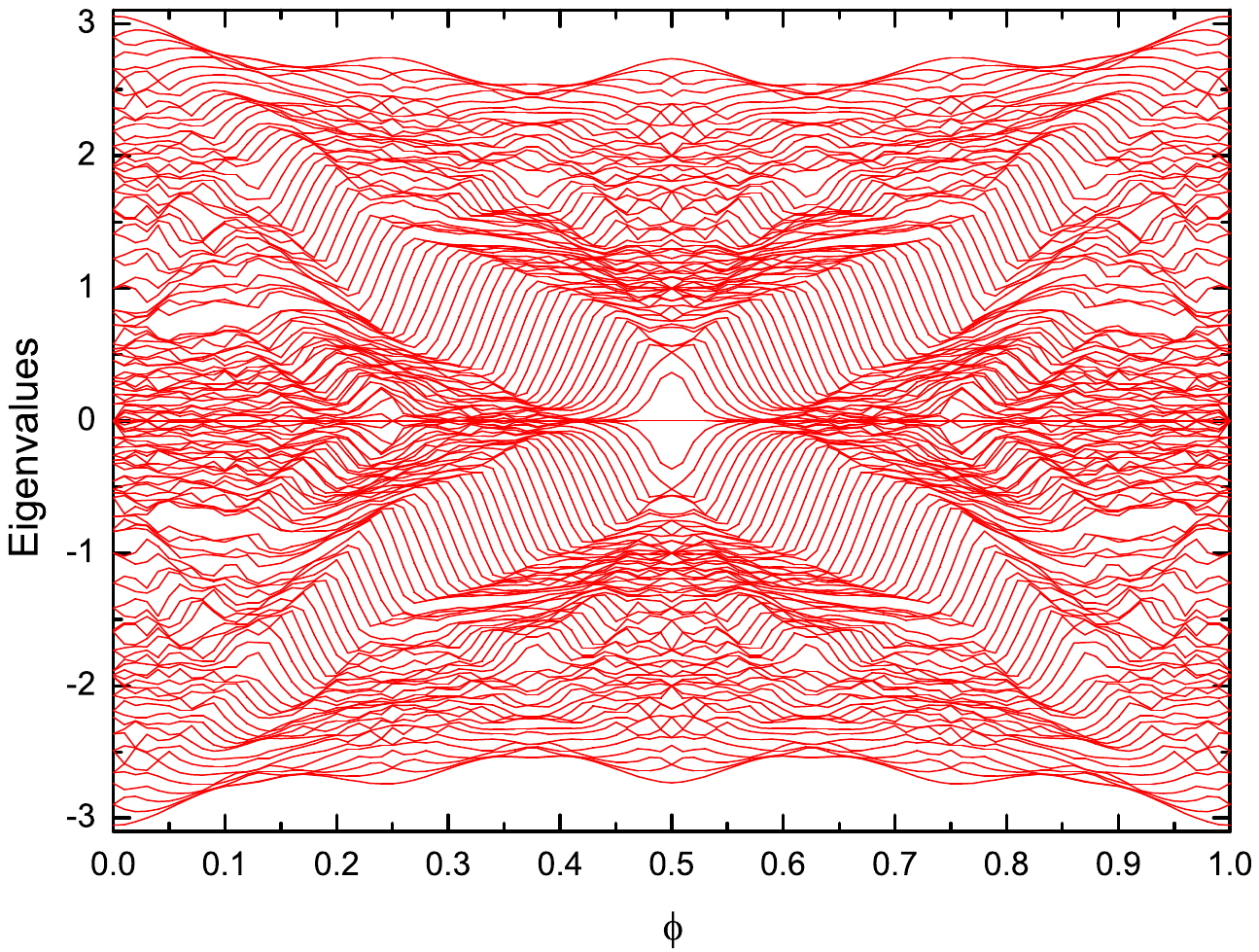}
\hskip-1.8cm
\includegraphics[angle=-00,width=0.4\textwidth]{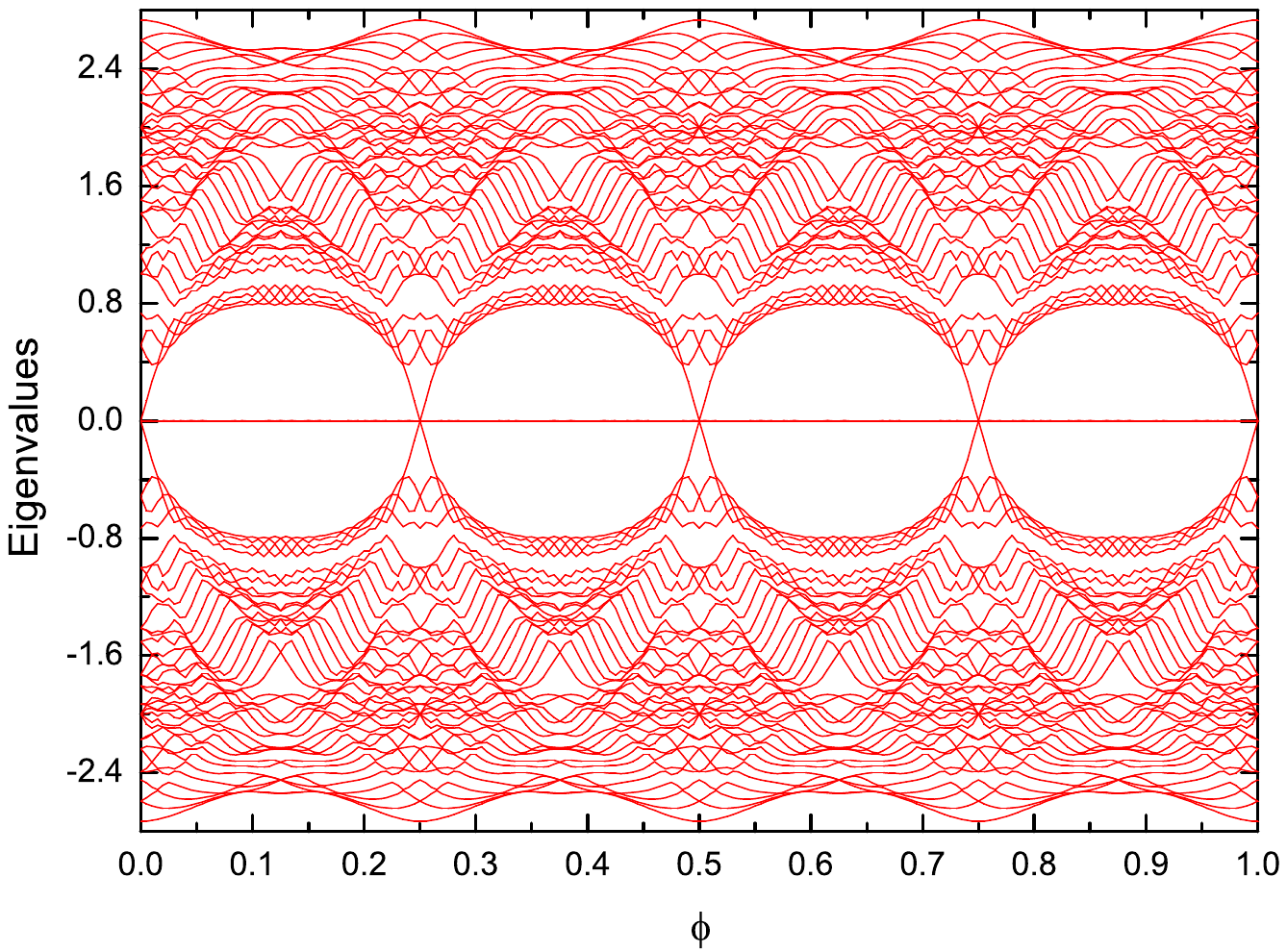}
\caption{ The energy spectrum as function of the  magnetic flux for
three values of the hopping integral $t'$ (see text) : (left) $t'=1$,
corresponding to the simple square lattice, (middle) $t'=0.5$, (right)
$t'=0$, corresponding to the square lattice with centered lines (Lieb
lattice). $\phi$ is the magnetic flux through the unit cell of the simple 
lattice  measured in quantum flux units. The Hofstadter butterfly is
obvious for $t'=1$,  while a doubled butterfly results for $t'=0$ in 
each of the intervals $\phi \in [0,0.25]$, $\phi \in [0.25,0.5]$, etc. 
(one has to keep in mind that the flux through the Lieb unit cell is 
four times larger than $\phi$). The energy is measured in units of hopping 
integral $t$.}
\end{figure}

\section{Specific aspects of the finite Lieb plaquette in magnetic field: zero energy flat band and twisted edge states}
\subsection{The properties of the flat band }
There are several pertinent questions which can be asked concerning the
flat band in the energy spectrum of the Lieb finite system: 
what is the degeneracy, 
what is the response to the magnetic and electric field and to the disorder?

Let us find first the conditions 
which should be satisfied by the zero energy eigenfunction $\Psi_{E=0}$.
Let $H$ be the  tight-binding Hamiltonian of a finite system 
and $\Psi_E$ its eigenfunctions:
\begin{equation}
 H=\sum_n E_n |n><n| + \sum_{n,m}t_{nm}|n><m|,~ ~ \Psi_E=\sum_n\alpha_n|n>,
\end{equation}
where $\{|n>\}$ is a  basis of functions localized at the sites $n$.
The condition $H\Psi_E=0$ generates a set of equations for the coefficients
 $\{\alpha_n\}$:
\begin{equation}
	E_n\alpha_n+\sum_m t_{nm}\alpha_m =0, ~~\forall n .
\end{equation}
 Eqs.(15) are  the necessary  
and sufficient  conditions  which must be fulfilled by the
wave function $\Psi_E$ in order to  correspond to the zero eigenvalue $E=0$.
With $E_n=0$, and taking into account only the nearest neighbors ($t_{nm}=t$)
the above equations become simply $\sum_{m\in {\mathcal V}_n} \alpha_m =0$, 
for any $n$, where the sum is taken over all  sites in the first vicinity 
${\mathcal V}_n$  of the site $n$.
In addition, if $\Psi^i_{E=0}$ and $\Psi^j_{E=0}$ are two degenerate states , 
the orthogonality condition reads $\sum_n\alpha^i_n\alpha^j_n =0$.
The number of configurations $\{\alpha_n\}$ which satisfy simultaneously the
two conditions equals the dimension of an orthogonal basis in the space of the
degenerate eigenfunctions at $E=0$.
\begin{figure}[htb]
\includegraphics[angle=-0,width=0.80\textwidth]{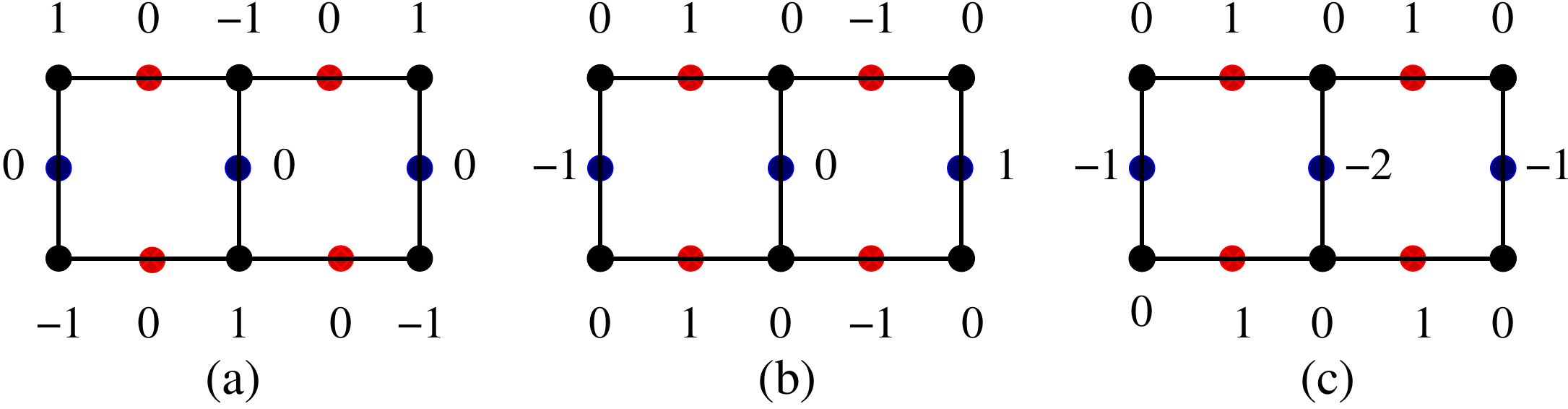}
\caption{(Color online) The three eigenstates of the flat band for a Lieb 
lattice composed of two cells.
The eigenfunctions are $\Psi^{(0)}=\sum_{nm}\alpha_{nm}|nm>$
and the coefficients $\alpha_{nm}$ are indicated. We notice that the condition
for the flat band appearance  
$\sum_{nm\in {\mathcal V}_{n_0m_0}}\alpha_{nm}=0$ 
holds for any site $\{n_0 m_0\}$.}
\end{figure}

An instructive illustration is the Lieb plaquette consisting of 
two cells (see Fig.4). The plaquette contains 13 atoms 
(6 of type $A$, 4  of type $B$ and 3 of type $C$). 
There are  three configurations of the coefficients $\alpha_n$ which satisfy
the conditions discussed above and they  are pictured as (a), (b) and (c).
(The numbers $\{0,-1,1,-2\}$ mentioned in Fig.4  represent the values, 
up to the normalization factor, of the coefficients $\alpha_n$).

With the notations used in the Hamiltonian (1), 
the three states can be written as:
\begin{eqnarray}
\nonumber \Psi^{(0)}_1(E=0,\phi=0)&=& [-a^{\dagger}_{11}+a^{\dagger}_{21}-
a^{\dagger}_{22}+a^{\dagger}_{12}-a^{\dagger}_{31}+a^{\dagger}_{32}]~|0> \\
\nonumber  \Psi^{(0)}_2(E=0,\phi=0)&=& [b^{\dagger}_{11}-b^{\dagger}_{21}
 -c^{\dagger}_{11}+c^{\dagger}_{31}
 +b^{\dagger}_{12}-b^{\dagger}_{22}]~|0> \\
\Psi^{(0)}_3(E=0,\phi=0)&=& [b^{\dagger}_{11}+b^{\dagger}_{21}
 -c^{\dagger}_{11}-2c^{\dagger}_{21}-c^{\dagger}_{31}
 +b^{\dagger}_{12}+b^{\dagger}_{22}]~|0>
\end{eqnarray}
It is obvious that  $\sum_n \alpha^i_n=0$ for any $i=1,2,3$ and that 
$<\Psi_{i}|\Psi_j>=0$ for any $i,j=1,2,3$, i.e. the three states correspond
to $E=0$ and are mutually orthogonal. 

Next, we want to find out how the zero energy states Eq.(16) respond to a 
perpendicular magnetic field.
In order to answer this question, we write the Hamiltonian (1) as:
\begin{eqnarray}
	H(\phi)&=&H^{(0)}(\phi=0)+H^{(1)}(\phi),\nonumber \\
	H^{(1)}(\phi)&=&\sum_{nm}\big(a^{\dagger}_{nm}b_{nm}+b^{\dagger}_{nm}
	a_{n+1,m}\big)(e^{-i\pi m\phi}-1) +H.c. ,
\end{eqnarray}
and  perform degenerate perturbation with respect to $H^{(1)}$. 
Applying this approach to the two-cell Lieb system the matrix elements 
involved are
 $<\Psi^{(0)}_1|H^1|\Psi^{(0)}_2>= 8i~sin\pi\phi$ and  
 $<\Psi^{(0)}_2|H^1|\Psi^{(0)}_3>=0$
 and the  secular equation reads:
\begin{displaymath}
          \det\left(\begin{array}{ccc}
          -E & 8i~ sin\pi\phi & 0 \\
          -8i~ sin\pi\phi &  -E & 0  \\
           0 & 0 & -E
	  \end{array}\right)
      =0 ,
\end{displaymath}
giving rise to the eigenvalues: 
$E_{1,2}=\pm 8t sin\pi\phi$ and $E_3=0$.

One remarks that the bulk state $\Psi_3$ does not couple to
the magnetic field and its eigenenergy remains $E_3=0$.
On the other hand, the surface states $\Psi_{1,2}$  get a dispersion 
which depends on $\phi$. The conclusion of the perturbative calculation 
is that the magnetic field reduces by 2 the degeneracy of the zero 
energy band.

Let us generalize now to a finite Lieb lattice 
 containing N cells along  the Ox-axis and M cells along Oy-axis, 
so that the total number of cells is $N_{cell}=NM$ and the number of  states 
is  3NM+2(N+M)+1.
It has been proved in the previous chapter that, at zero magnetic field,
the number of  zero energy degenerate states is $N_{cell}+1$.
Then, the two-cell model shows that in the presence of the magnetic field
two states separates from the bunch so that the degeneracy of the 
flat band becomes $N_{cell}-1$.
Using a similar approach for the general case, one has to use the 
eigenfunctions Eq.(11) and Eq.(13) and the expression Eq.(17) as the
perturbation. One finds out easily that 
$<\Phi^0(\vec{k})|H^{(1)}\Phi^0(\vec{k'})>=0$,
and  that the only nonvanishing matrix elements are $X(\vec{k})=:
<\Phi^0(\vec{k})|H^{(1)} \Phi^0_a>$.
In the general case, the secular equation becomes:
\begin{displaymath}
    \det\left(\begin{array}{ccccc}
    -E &0 &~ 0 & . . .& X(\vec{k}_1) \\
     0 &-E &~ 0 & . . .& X(\vec{k}_2)\\
    . . .&. . .& . . .& . . .&. . .  \\
    X(\vec{k}_1) & X(\vec{k}_2) &~ X_3&~ ~. . . & -E \\
\end{array}\right)
       =0 ,
\end{displaymath}
which in the polynomial form reads
$ E^{N-2} (E^2-X^2) =0 $, where $X^2=X^2(\vec{k}_1)+. . . X^2(\vec{k}_{N-1})$.
This formula (where $N$ stands here for the degeneracy of the flat band) 
says that from the whole bunch only two levels get a dispersion depending 
on $\phi$, meaning that the degeneracy of the zero energy level is 
reduced by $2$ in the presence of the magnetic field.
So,  the general finite Lieb plaquette behaves similarly
to the two-cell model.

The numerically calculated energy spectrum of the finite plaquette
in perpendicular magnetic field
is shown in Fig.5a, where one can check again the presence of 
the two levels which separates from the flat band while the most of the bunch
at $E=0$  consisting of  $N_{cell}-1$ states remains dispersionless.
\begin{figure}[htb]
\includegraphics[angle=-0,width=0.45\textwidth]{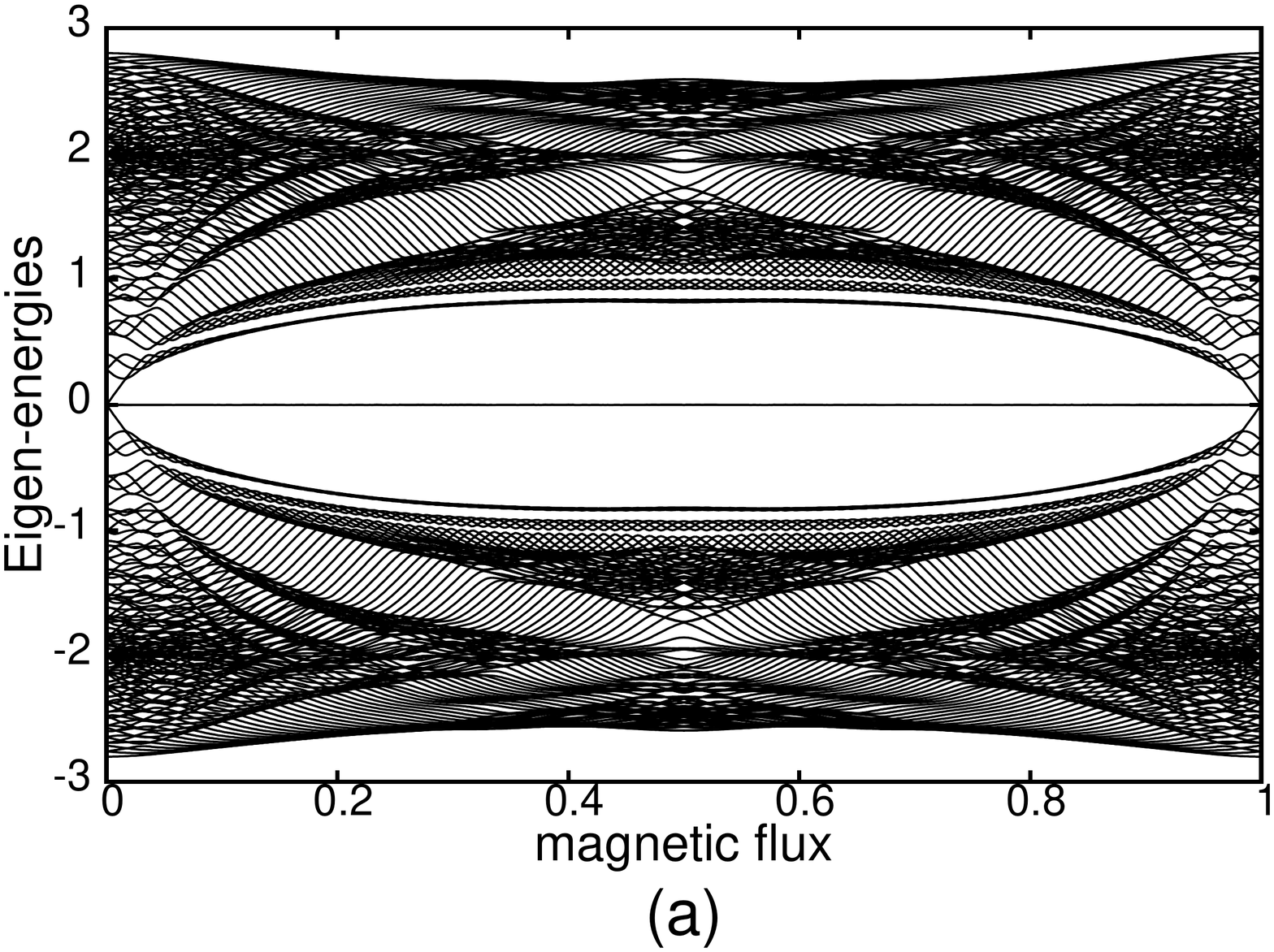}
\includegraphics[angle=-0,width=0.45\textwidth]{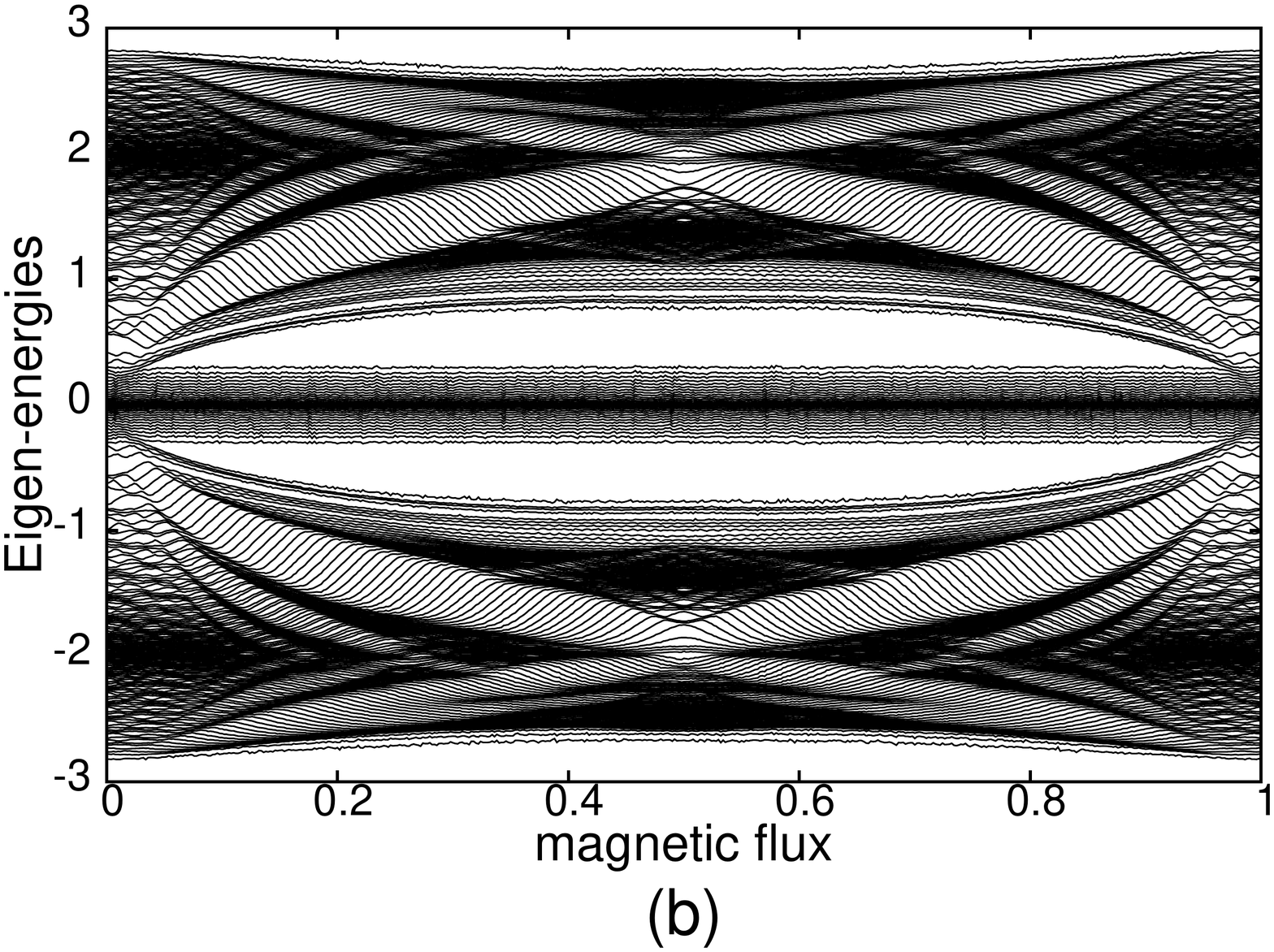}
\vskip0.2cm
\caption{The Hofstadter-type spectrum of a finite Lieb lattice of 
dimension $N^x_{cell}=N^y_{cell}=10$. (a) for  the clean plaquette, and (b)
for the disordered one (disorder strength $W=1$); the flux $\phi$ is measured 
in quantum flux units.}
\end{figure}

The strong degeneracy of the flat band can be however lifted by a disordered
potential. The broadening of the band depends on the strength of the disorder,
however it remains independent of the magnetic field  as
in the case of the clean system (see  Fig.5b).
We use a diagonal  disorder of the Anderson-type
characterized by the width parameter $W$ \cite{note1}.
The calculation of both the inverse participation number (IPN) and of the
 interlevel distribution  indicate that in the middle of the 
disordered band  the states are still delocalized, 
and described  by the orthogonal Wigner-Dyson distribution
($\beta=1$)  which  is the typical result in the absence of the magnetic field. 
This proves once more the absence of response of the flat band   
to the perpendicularly applied magnetic field, even in the presence 
of disorder. 

The inverse participation number (IPN) is defined as: 
\begin{equation}
IPN_E=\sum_n |<n|\Psi_E>|^4
\end{equation}
and indicates the degree of localization of the states. The small values 
of the IPN for energies in the middle of the density of states denotes 
the presence of extended states, and, as expected, the 
localization increases towards the band edges. The numerically calculated
density of states and the dependence on energy of the inverse
participation number are shown in Fig.6a. 
Further information about
the localization and the response to the magnetic field is provided by the
distribution function of the level spacing between consecutive eigenvalues 
 $s_n=E_n-E_{n-1}$ of the disordered system. Let us define  the
dimensionless quantity 
$t_n=s_n/<s_n>$, where $<s_n>$ is the mean level spacing.
In the disordered system, in the range of delocalized  states, 
the level spacings are distributed according to the Wigner-Dyson surmise 
\cite{Mehta}:
\begin{equation}
\mathcal P (t)=b_{\beta}t^{\beta}e^{-a_{\beta}t^2},   
\end{equation}
where $\beta=2$ in the  presence of the magnetic field,  
and $\beta=1$  if $B= 0$. 
As a signature of the distribution, the variance of the level spacing   
$\delta t=<\delta s>/<s>$  is $<\delta t>=0.4220$ in the first 
case, and $<\delta t>=0.5227$ in the second one. 
Fig.6b shows the numerically calculated variance of the level spacing
distribution, and one can notice that, in the middle of the flat band, 
where the states are delocalized, the variance is $<\delta t>=0.5227$. 
This means that, despite the presence of the magnetic field, the flat band
behaves according to  the orthogonal ($\beta=1$)
Wigner-Dyson distribution instead of the unitary one ($\beta=2$), as it is
expected at $B\ne 0$. 

\begin{figure}[htb]
\hskip-1.7cm
\hskip1.5cm
\includegraphics[angle=-0,width=0.45\textwidth]{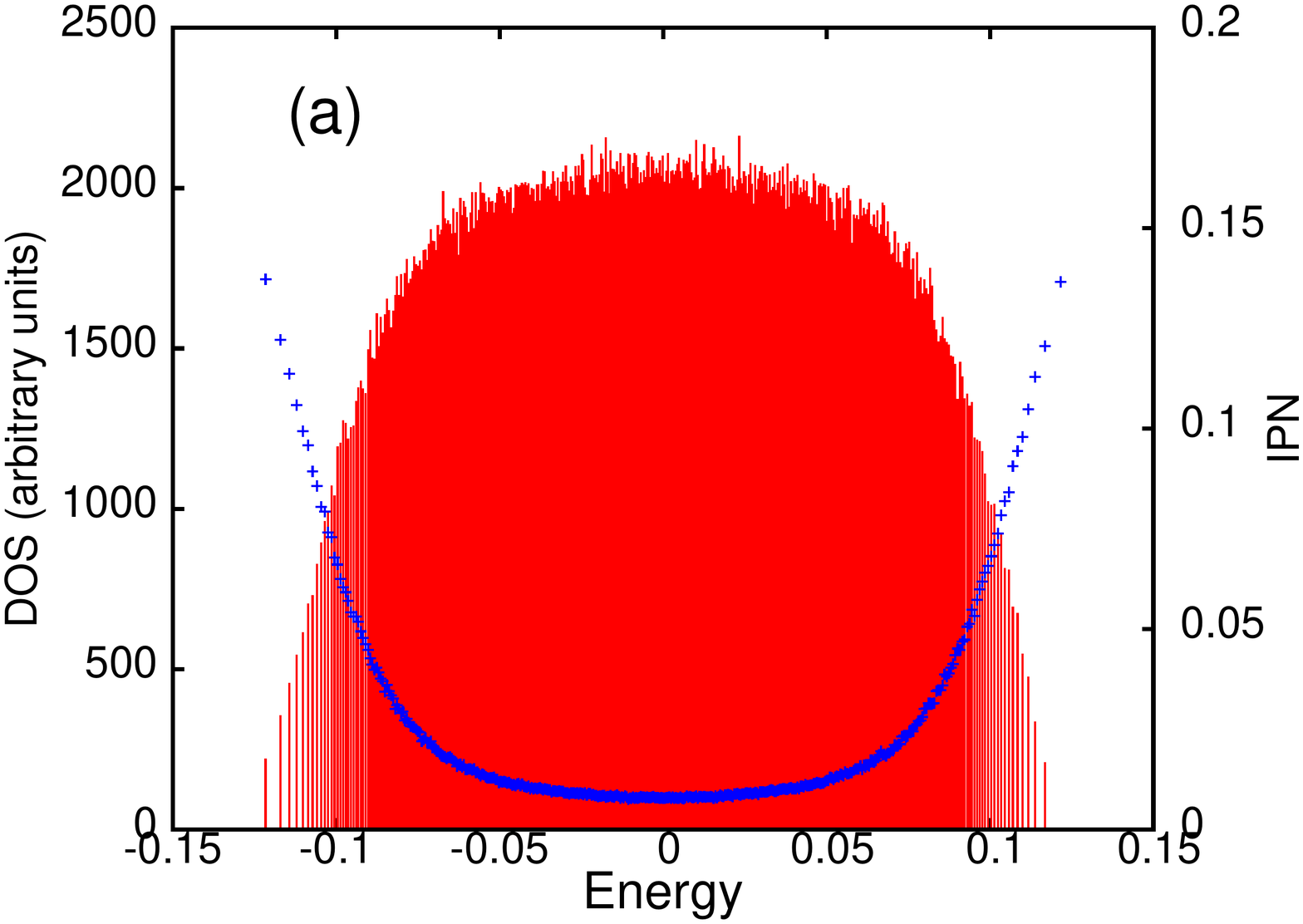}
\hskip0.70cm
\includegraphics[angle=-0,width=0.45\textwidth]{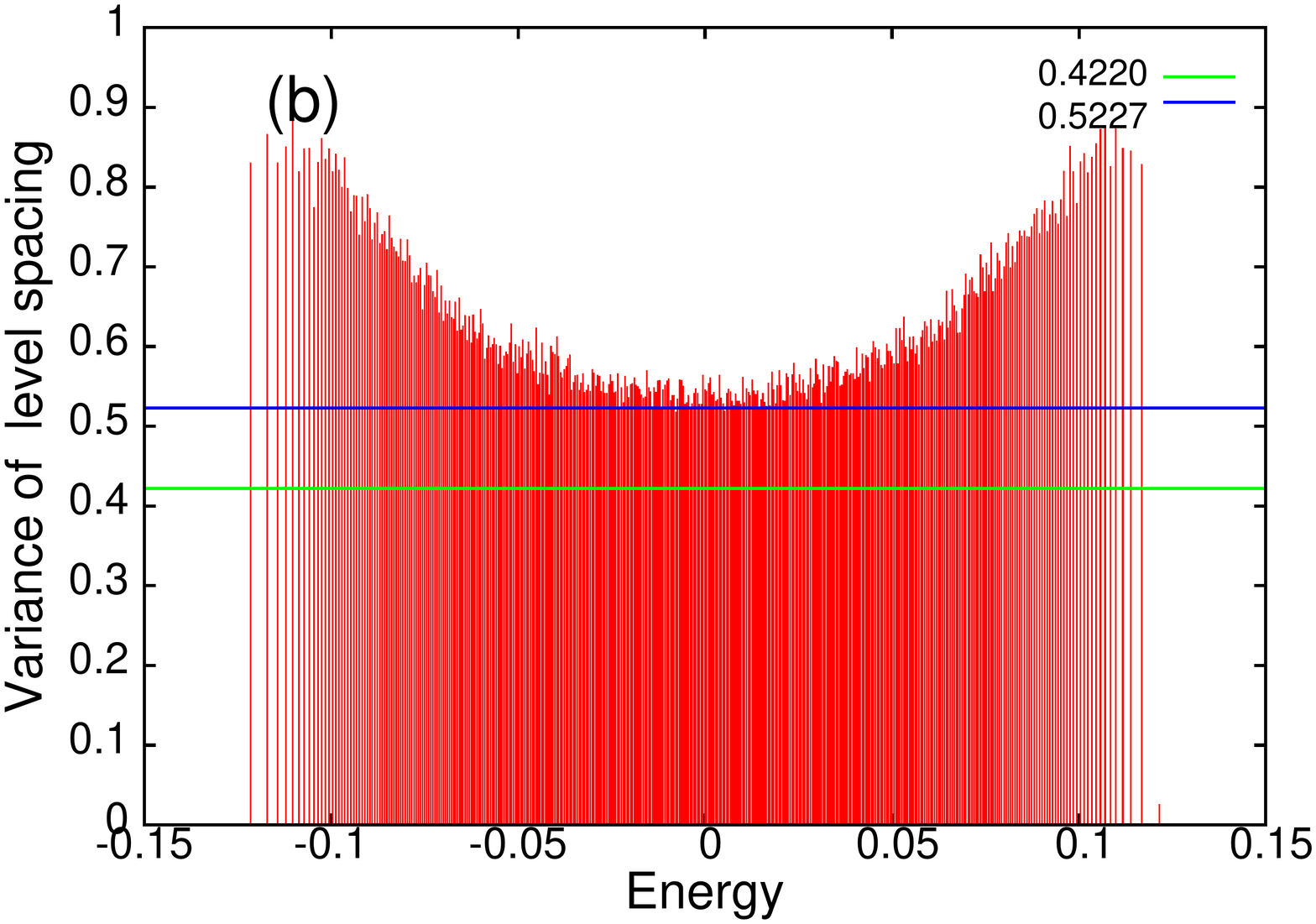}
\caption{(Color online)(a) The density of states 
and IPN in the flat band range as function of the energy for a disordered
Lieb plaquette of dimension $20\times20$ cells  averaged over 
1000 configurations (disorder strength $W=0.3$). 
(b) The variance of the level spacing distribution
as function of energy for the same disordered system; the horizontal lines
correspond to 0.4220 (as for the unitary ensemble) and 0.5227 
(as for the orthogonal ensemble). }
\end{figure}

Another way to lift the degeneracy of the zero-energy band is to apply
an in-plane static electric field. We expect specific aspects coming from 
the existence of the edges  and of the lattice structure.
In the numerical calculation the electric field applied along Oy-axis is 
simulated by replacing the atomic energies 
$E_{nm}$ with $E_{nm}+{\mathcal E} y_n   $,
where $y_n$ is the  site coordinate along Oy.
Fig.7a  shows how the eigenvalues stemming from the flat band are split
in several degenerate mini-bands which develop a Stark fan with increasing
electric field. It can be checked that the number of mini-bands equals the
number of lattice cells along the direction of the electric field.
A perpendicular magnetic field gives rise to supplementary fine
splitting and to the presence of states  between
mini-bands. This can be seen in Fig.7b and also in Fig.8.
We have noticed   that the flat band states are much more 
sensitive to the electric field than the edge states, and  they give rise to a 
Wannier-Stark ladder at values of the electric field $\mathcal E$ for which 
the edge states are still non affected.  We have also numerically observed 
that the wave function  in the $l-th$ miniband   is mainly localized in the 
$l-th$ row of cells  in the direction of the electric field.
\begin{figure}[htb]
\includegraphics[angle=-00,width=0.45\textwidth]{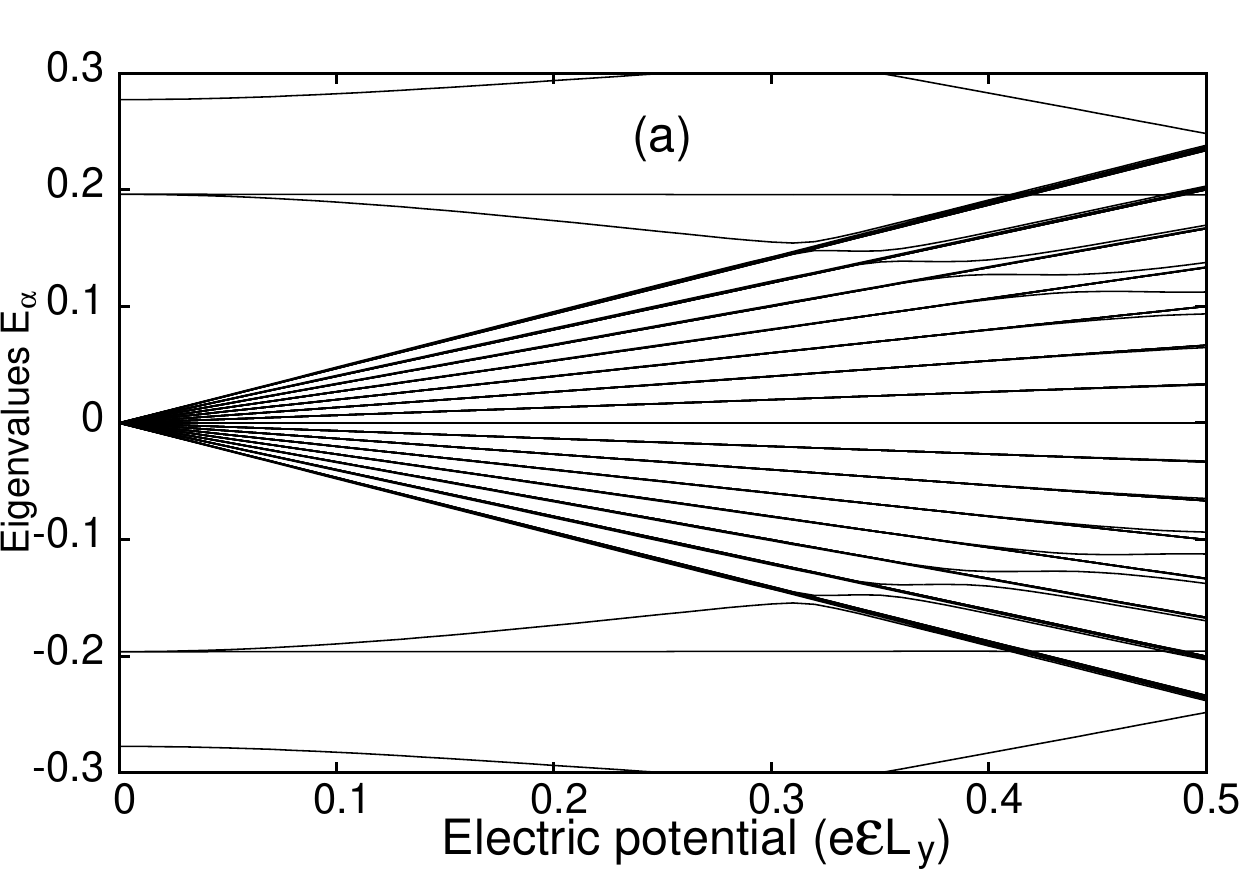}
\includegraphics[angle=-00,width=0.45\textwidth]{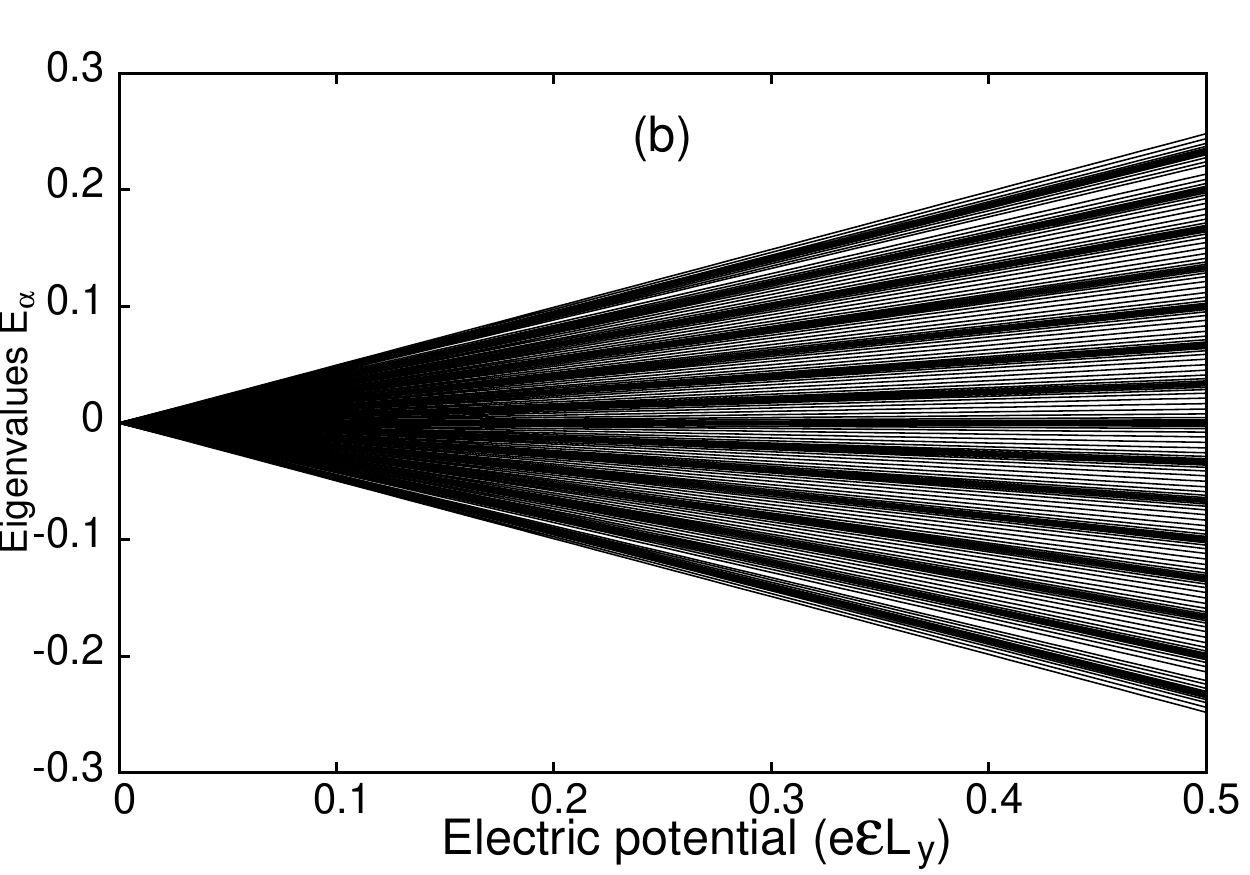}
\caption{The low energy spectrum of a finite Lieb plaquette as function
of the electric potential applied on the plaquette in the Oy-direction at: 
(a) $\phi=0$, and (b) $\phi=0.12$.}
\end{figure}

We already have seen that the flat band states do not show any diamagnetic 
response, and it is somehow surprising that  the Wannier-Stark states 
coming from the former flat band exhibit a diamgnetic moment when the
magnetic field is applied. It is interesting that each mini-band shows both
positive and negative magnetic moments, and Fig.8a  suggests 
that the  chirality $dE/d\phi$  changes the sign at the center of the
mini-band.
\begin{figure}[htb]
\includegraphics[angle=-00,width=0.45\textwidth]{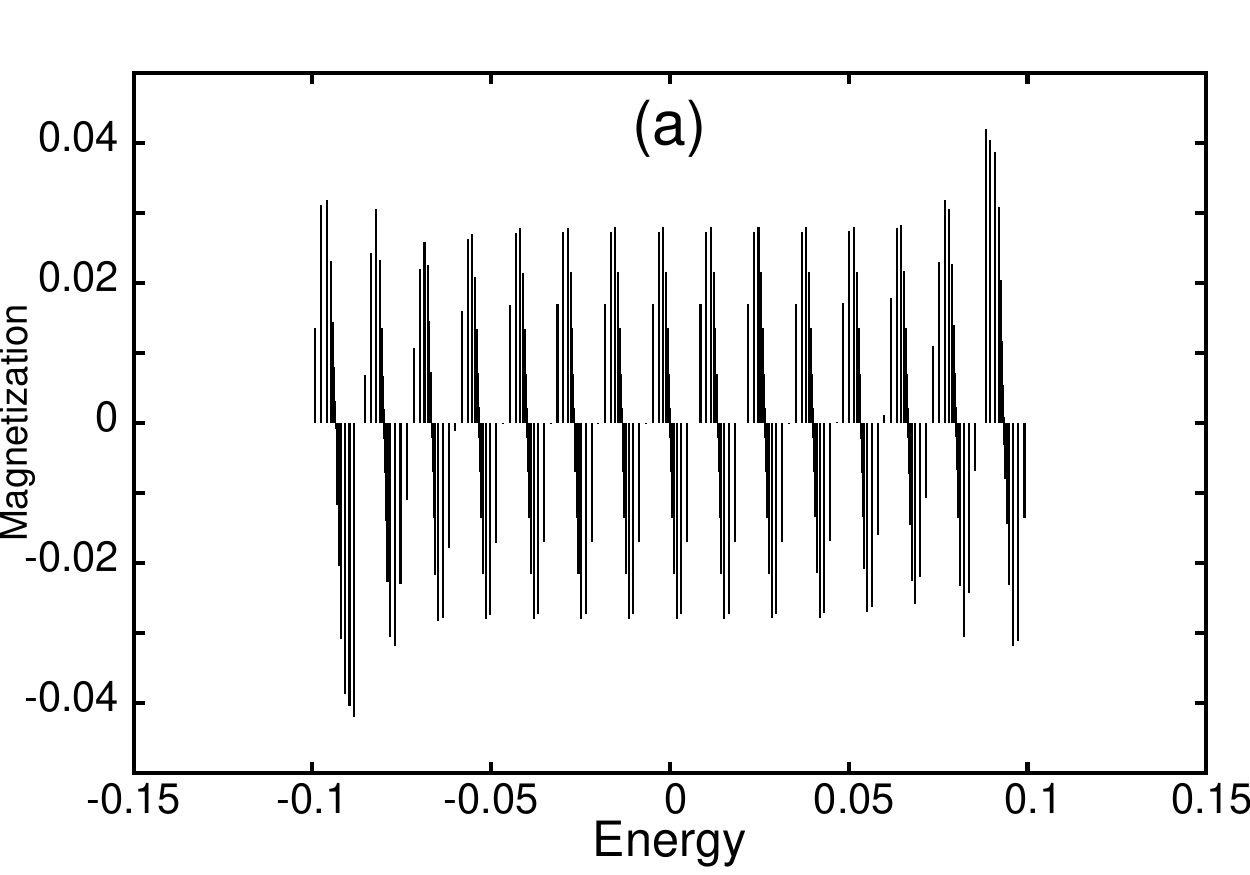}
\hskip0.2cm
\includegraphics[angle=-00,width=0.45\textwidth]{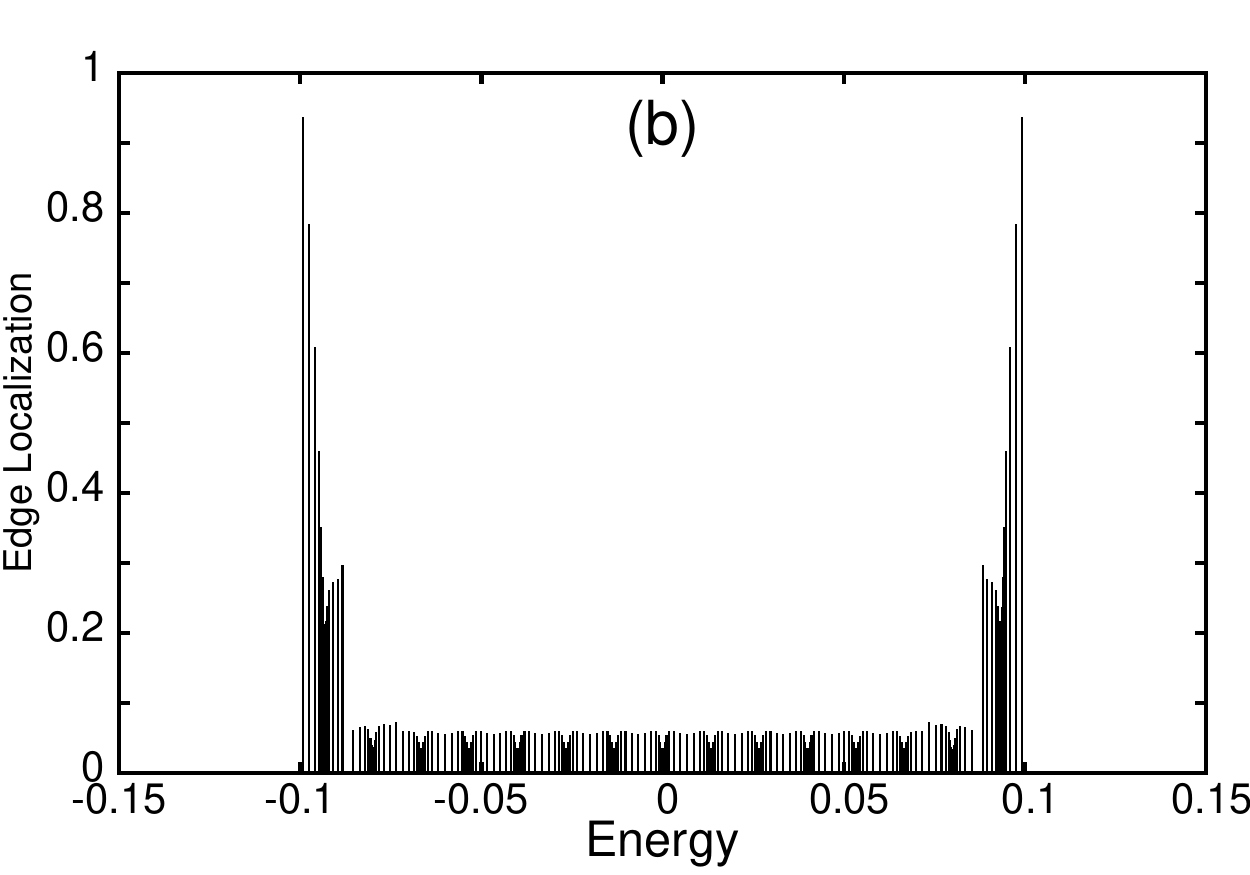}
\caption{The behavior of the flat band in crossed magnetic and electric field.
(a) The orbital magnetization $M_\alpha$,  and  (b) the edge 
localization $P_\alpha^{edge}$ (b) vs. energy $E_\alpha$ 
for a finite Lieb lattice of dimension $N^x_{cell}=N^y_{cell}=15$.
The flat band turns into a set of 15 minibands, 
every miniband being composed of two parts with opposite magnetization.
The states in the lowest and highest miniband have significantly increased 
edge localization ($\phi=0.12$  and $e \mathcal {E} L_y=0.2$).}
\end{figure}
We have studied also the localization properties of the eigenstates,
particularly the localization along the edges $P^{edge}_{\alpha}$, defined as:
\begin{equation}
	P^{Edge}_{\alpha}=\sum_{i\in Edge}|\Phi_{\alpha}(i)|^2,
\end{equation}
where the index $\alpha$ indicates the state, and the sum is taken over
all the sites which belong to the plaquette boundary. It turns out  
that the states which belong to the mini-bands from the extremities of the
fan spectrum are strongly localized along the edges (Fig.8b). The localization
is of electric origin since the picture is similar no matter whether the
magnetic field is present or not.

We conclude, saying that the disorder lifts the degeneracy of the flat band
keeping the states independent of the magnetic field, while the electric field
produces states which  respond to the magnetic field 
and show  specific diamagnetic moments.

\subsection{The  twisted and type II edge states and their properties}
The confinement of  the Lieb lattice  induces several types of edge states.
Besides the conventional edge states found in the 
Bloch-Landau and Dirac-Landau regions,
there are still two other classes of  edge states. 
We discuss first the {\it twisted} edge states lying in the magnetic gap 
opened around the  degenerated energy level $E=0$.
Although the new  states are localized along the perimeter of the plaquette,
they do not follow the known behavior of the conventional edge  states.
The new class of edge states manifest specific properties:
i) their energy depends on the flux in a periodic way. This means that the
chirality defined by the sign of $dE/d\phi$ is not conserved but alternate when
changing the flux, in contradistinction  to the usual edge states either in
the Bloch-Landau or  Dirac-Landau domain. Obviously, the alternate 
chirality  should be reflected also in oscillations  of the  
orbital magnetization  at the variation of the  magnetic flux. 
ii) their energies as function of the flux 
appear as twisted into bunches; for the clean square plaquette 
shown in Fig.9a  each bunch consists of four states.
iii) the states prove the lack of robustness against disorder and 
iv) prove specific transport properties, namely, the twisted edge states carry a finite longitudinal resistance accompanied by vanishing Hall resistance.
\begin{figure}[htb]
\includegraphics[angle=-0,width=0.45\textwidth]{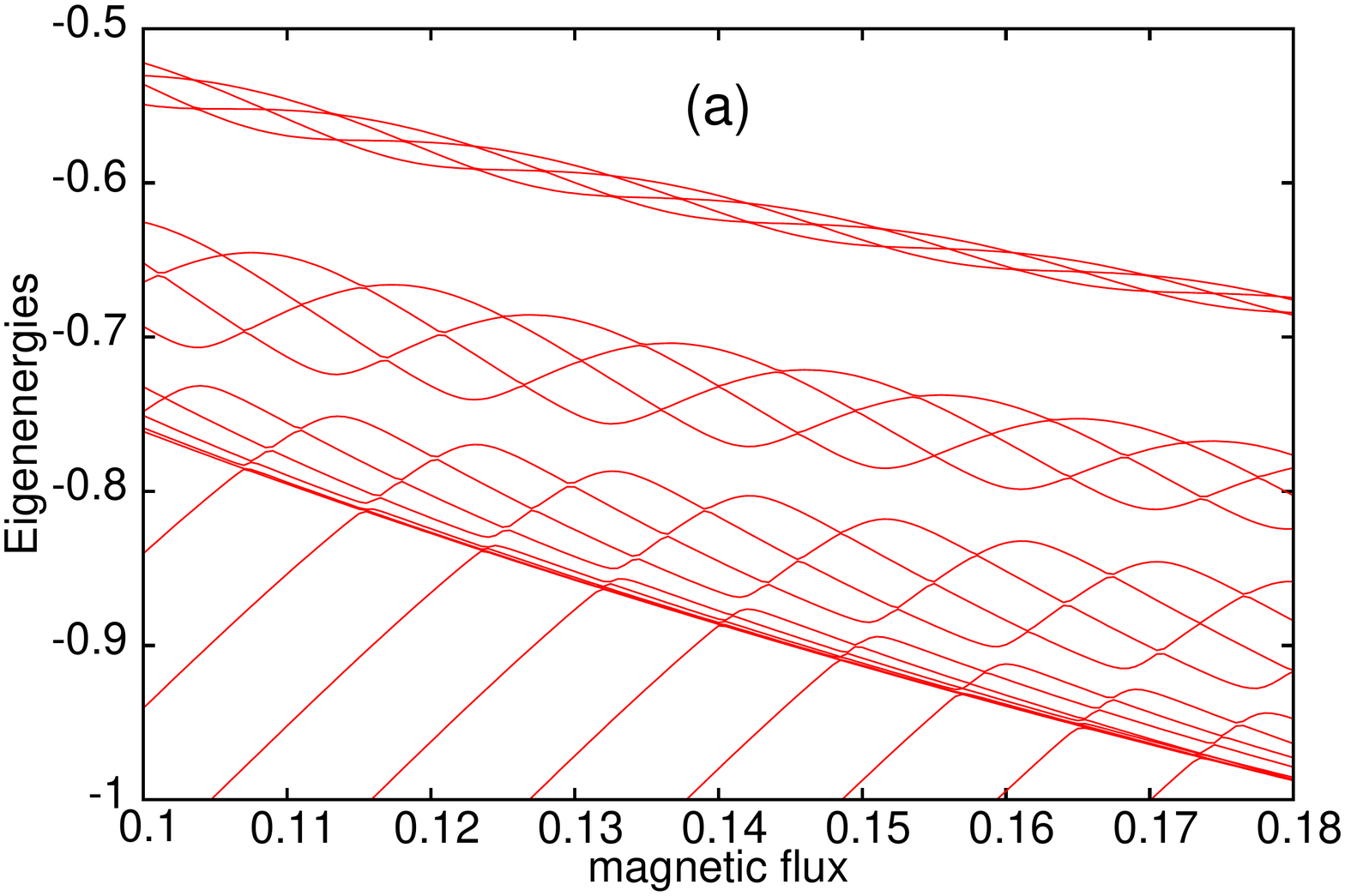}
\includegraphics[angle=-0,width=0.45\textwidth]{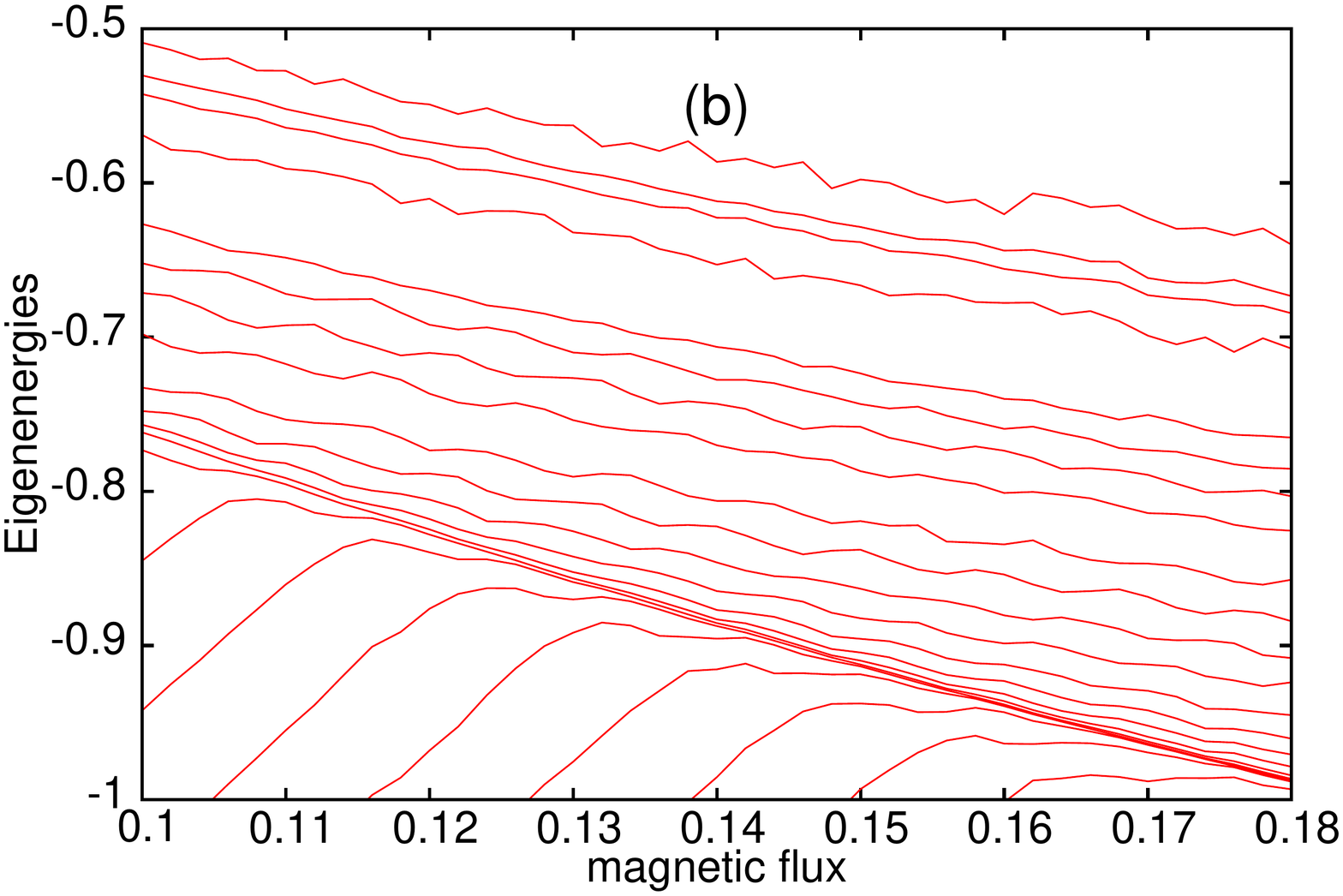}
\caption{The eigenenergy in the range of the twisted edge states vs.
the magnetic flux $\phi$
for a pure Lieb lattice (a) and for a disorder Lieb lattice (b).
The twisted edge states has an oscillatory behavior when the magnetic flux
is varied, and they form bunches with four states in each bunch.
The oscillatory behavior is destroyed by disorder in the right figure,
but the conventional edge states  (shown in the lower part
of the spectrum) remain robust against disorder.
The dimension of the Lieb lattice is $N^x_{cell}=N^y_{cell}=10$
and the amplitude of the Anderson disorder is $W$=1.}
\end{figure}
A piece of the spectrum of the clean plaquette in the energy range of  
twisted edge states is shown in Fig.9a, where
bunches consisting of four twisted edge states can be observed.
One also has to notice that, at a given flux, the 
states in the bunch may show opposite chirality meaning  that they 
carry diamagnetic currents moving in opposite directions. 
In the presence of disorder (Fig.9b) one notices that the twisted 
eigenenergies get stretched but the rest of the spectrum 
(the band and the edge states in the Dirac region) is not affected. 
This indicates that the twisted states are very sensitive to disorder.
The understanding of this effect is simple in the sense that the 
degeneracy at  crossing  points \cite{note2} is lifted by the  
perturbation introduced by the impurity  potential, and 
this  occurs even at weak disorder. 
\begin{figure}[htb]
\includegraphics[angle=-0,width=0.40\textwidth]{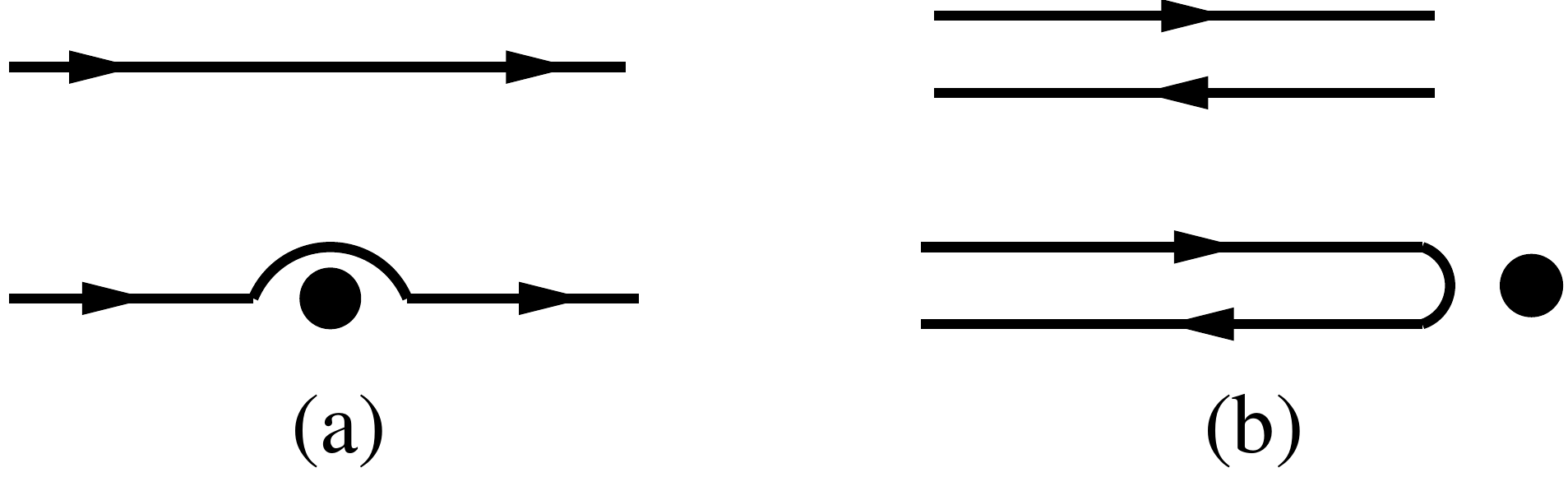}
\vskip-0.5cm
\caption{The behavior of the edge states with disorder.
(a) The absence of the backscattering for a conventional edge states.
A pair of twisted states which are sufficiently close in energy 
may suffer the backscattering suggested in (b), which induces the 
localization shown in Fig.11.}

\end{figure}
\begin{figure}[htb]
\includegraphics[angle=5,width=0.45\textwidth]{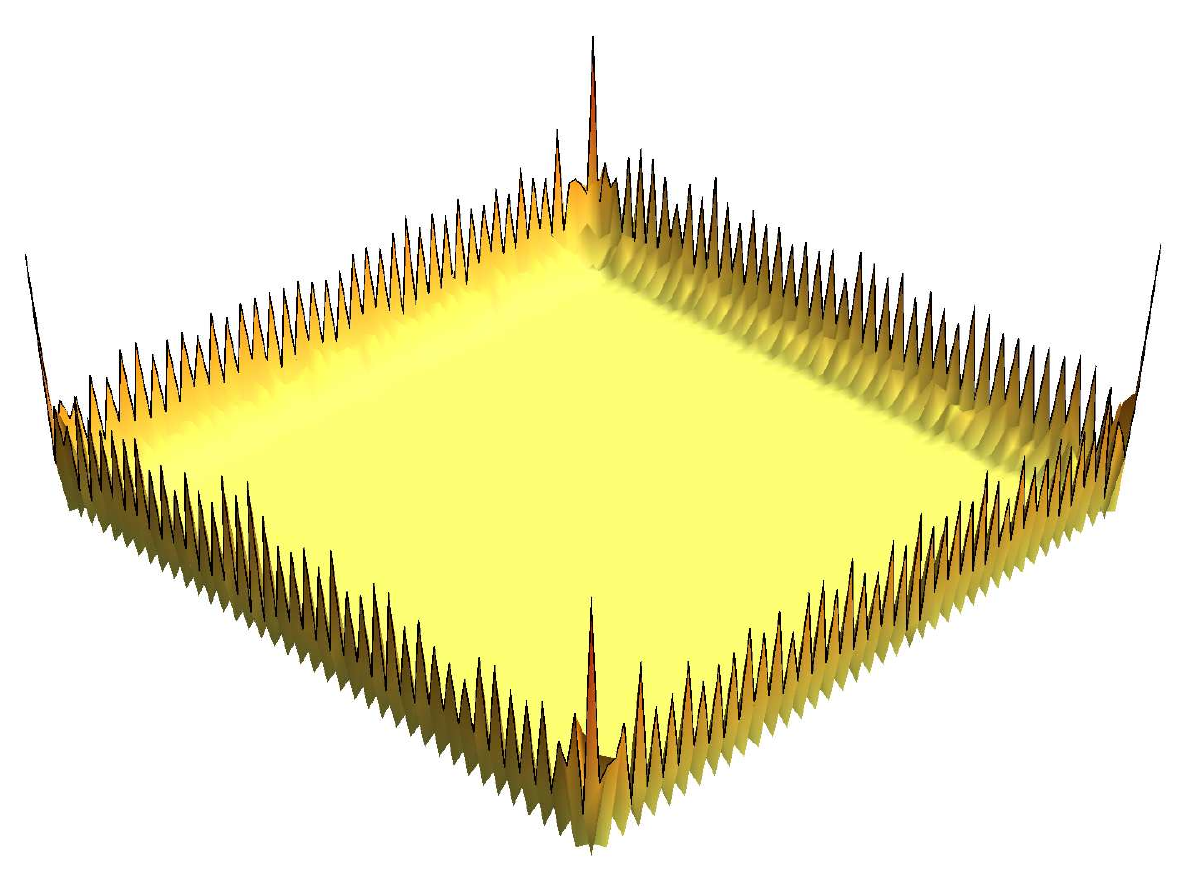}
\includegraphics[angle=10,width=0.4\textwidth]{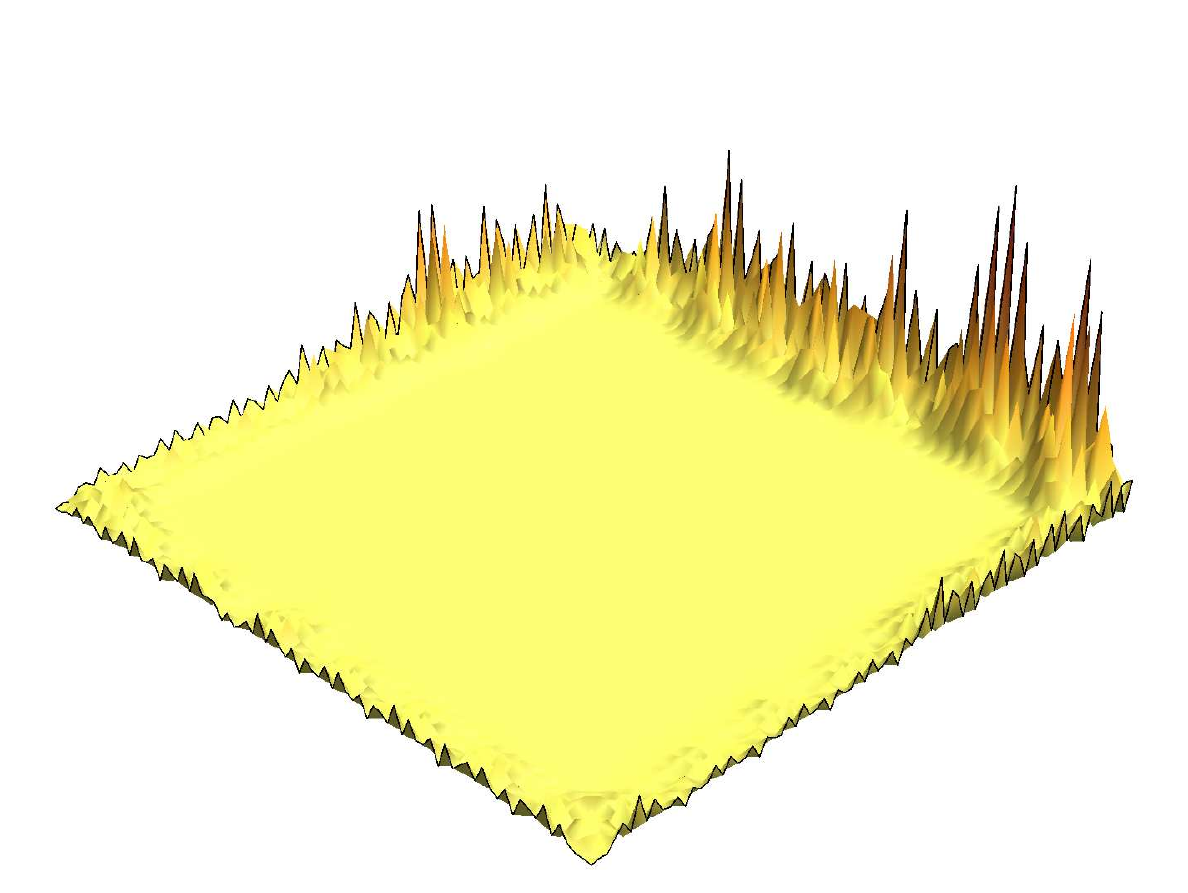}
\caption{ $|\Psi|^2$ calculated for 
 a conventional edge state in the Dirac range (left)
 and for a twisted edge state  (right)
for a disordered plaquette with  $W=0.2$.
One observes that this low disorder does not
affect the  conventional edge state but localizes the twisted  state.}
\end{figure}
\begin{figure}[htb]
\includegraphics[angle=-0,width=0.5\textwidth]{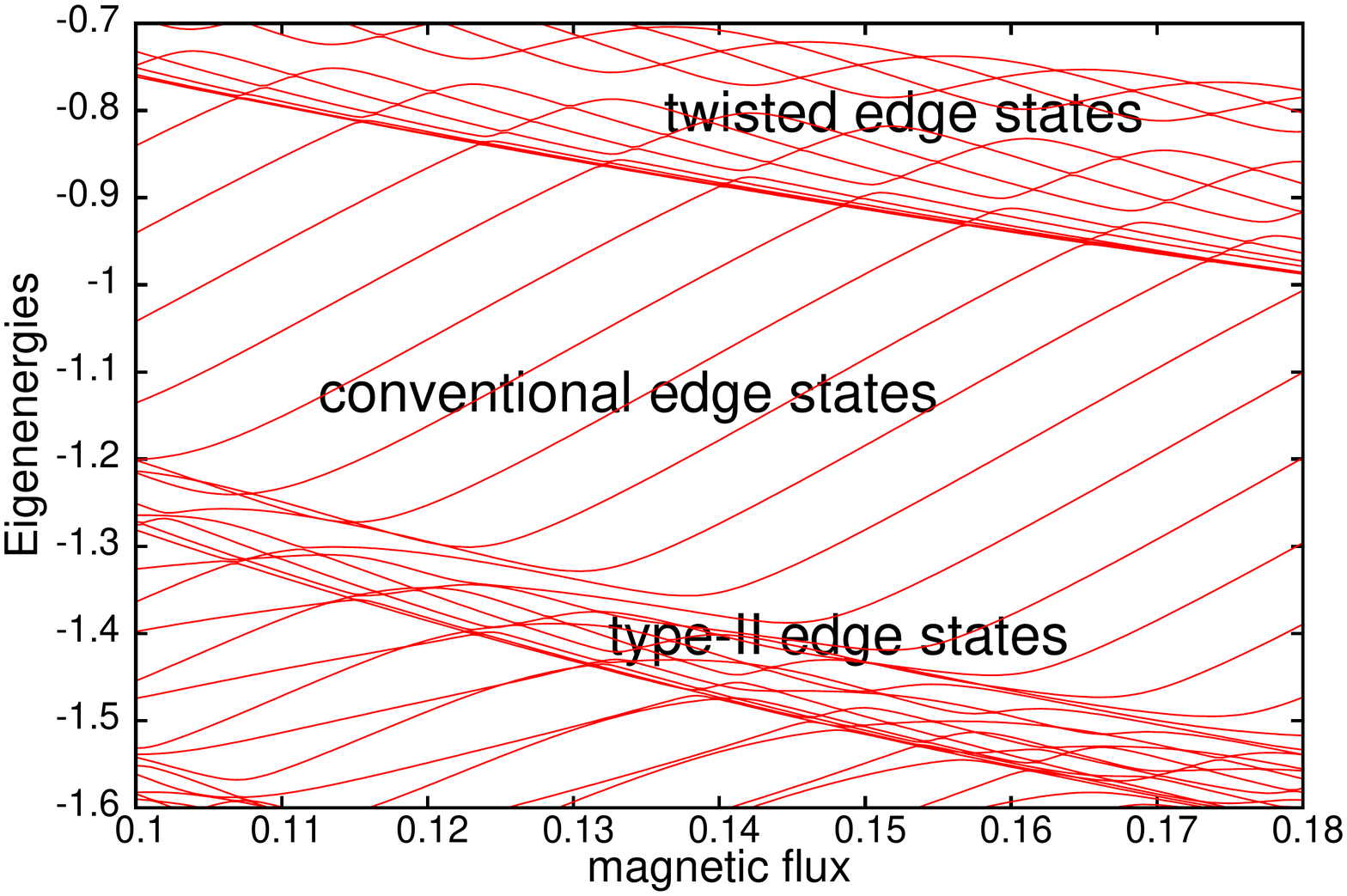}
\caption{
The eigenenergies vs. the magnetic flux $\phi$ for a pure Lieb lattice
in the range which emphasizes the type II edge states. In the spectrum, 
they are located  between the  bulk states in the  Dirac-Landau range and 
the conventional edge states of the first gap. 
Their energy decreases with the magnetic field similar to  the bulk states,
however they have edge localization of the wave function. 
The dimension of the Lieb lattice is $N^x_{cell}=N^y_{cell}=10$.}
\end{figure}
The Lieb lattice exhibits still another specific edge states (which we 
call type-II edge states), which in Fig.12  are placed immediately above 
the Dirac-Landau bands at the transition from Dirac bulk to conventional
edge states. They cannot be identified according to the sign 
of the magnetic moment \cite{Nita} since their chirality $dE_n/d\phi$ 
is the same as  for the bulk  (band) states  \cite{note3}.
\begin{figure}[htb]
\includegraphics[angle=-00,width=0.5\textwidth]{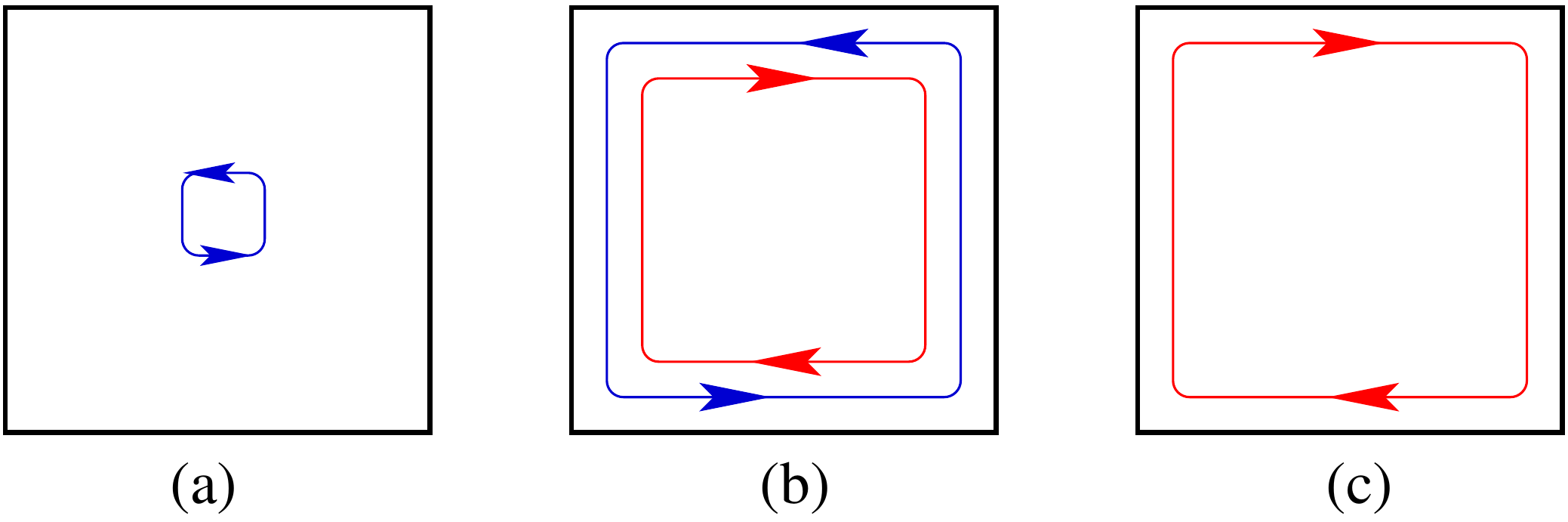}
\caption{
The sketch of the diamagnetic currents in the Dirac-Landau range of the
spectrum: (a) the counterclockwise loop of a bulk state, 
(b) the double ridge current of a type-II edge states, 
and (c) the clockwise loop of a conventional edge states.
The twisted edge states may show both (b) and (c) aspect.}
\end{figure}
Nevertheless, the  diamagnetic currents of these states are
located along the edges of the plaquette. 
These edge states show a double-ridge profile and 
carry current in both directions, but nonetheless  the total magnetization
remains of bulk-type.

In Fig.13 the diamagnetic currents of bulk states, type-II edge states 
and of  conventional edge states are sketched. The twisted edge states 
may show currents similar to both conventional and type-II edge states.
Compared to the twisted states, the type-II edge states behave
substantially different in the electronic transport. 
These states will be studied in detail elsewhere.
\begin{figure}[htm]
	\includegraphics[angle=-00,width=1.2\textwidth]{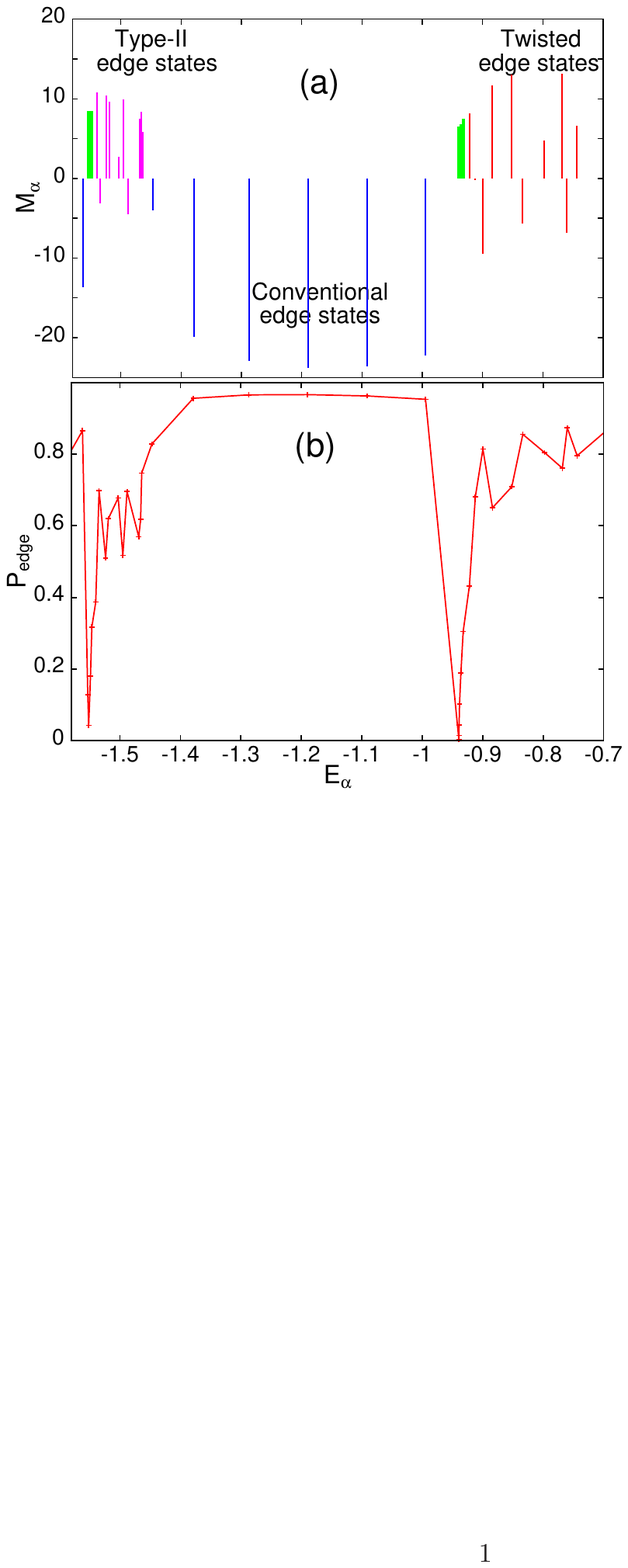}
\vskip-14cm
	\caption{(Color online)  Magnetization $M_{\alpha}$ and 
	localization at the edges $P_{edge}$ corresponding to 
	the eigenenergies $E_{\alpha}$: the bulk (band)  states (green)
	show positive magnetization and vanishing localization at the edges;
	the conventional edge states (blue) show negative magnetization and
	$97\%$ localization at the edges; the type-II and the twisted states
	show $60-80\%$ localization at the edges.
	The data are for a clean Lieb plaquette of dimension $N_{cell}^x=
	N_{cell}^y=10$ and $\phi=0.16$.}
\end{figure}
The contribution to the magnetization of each eigenstates 
$|\alpha>$ is calculated following the approach from \cite{Aldea-magn}, namely:
 \begin{eqnarray}
M_{\alpha}=&&- <\alpha|\frac{dH}{d\phi}|\alpha>\nonumber \\
=&&i\pi t_x \sum_{mn} \big(me^{-i m\pi\phi} <\alpha|A_{nm}><B_{nm}|\alpha>
        -me^{i m\pi \phi} <\alpha|A_{nm}><B_{n-1,m}|\alpha>\nonumber\\
         +&&me^{-i m\pi\phi} <\alpha|B_{nm}><A_{n+1,m}|\alpha>
        -me^{i m\pi\phi}  <\alpha|B_{nm}><A_{nm}|\alpha>\big).
\end{eqnarray}
All the matrix elements  in the above equation are known once the eigenstates
$|\alpha> $ are calculated numerically in the presence of the magnetic flux.
Fig.14 depicts the diamagnetic moments and the localization at the edges of 
different type of states. The bulk (band)  states
 show positive magnetization and vanishing localization at the edge,
the conventional edge states  show negative magnetization, and
 $97\%$ localization at the edge. The twisted edge states show alternating
 magnetization, as expected, but also an unanticipated differences in the
 degree of edge localization. This is because they exhibit
 either a single- or double-ridge wave function. Obviously, the double-ridge
 wave function is not  strictly stuck to the edge so that $P_{edge}\approx
 0.7-0.8$, while for the single-ridge states the same parameter goes up
 to $0.9$. Fig.14 points out that the single-ridge states which are localized 
 close to the edge exhibits  negative magnetization (as the conventional edge
 states), while the double-ridge states exhibits positive magnetization.

\section{The integer quantum Hall effect}
The quantum transport of the 2D Lieb plaquette shows some similarities with 
the case of graphene, however it also reveals particular properties. 
The Hall resistance  as function of the Fermi energy at a given
quantizing magnetic field was obtained in Ref.3  by calculating
the Chern numbers, and has the general aspect which can be observed 
in  Fig.15 which we obtain in the Landauer-B\"{u}ttiker formalism:
starting from the  bottom of the spectrum,
$R_H$ shows $h/e^2$ steps in the Bloch-Landau region, then
change the sign, and shows again $h/e^2$ steps in the Dirac-Landau 
region. The steps of the quantum Hall plateaus 
differs from those of the  honeycomb lattice since in the Lieb case there 
is only one Dirac cone per the unit cell. The change of sign is associated 
with  the opposite chirality of the edge states in the two regions and 
occurs at $E=\pm 2t$, while in graphene the same change takes place 
at $E=\pm t$ \cite{graph}. 
The density of states (shown in blue in Fig.15)
is calculated as $DOS=-\frac{1}{\pi}\sum_n Im G^+_{nn}(E)$, where $G^+$ is the 
retarded Green function for the mesoscopic plaquette connected to the leads.
In order to calculate the transport properties the mesoscopic plaquette must be
connected to leads, the whole system being described in the  tight-binding 
approach by the Hamiltonian:
\begin{equation}
H=H^S+H^L+\tau H^{LS},
\end{equation}
where the first term is just the Hamiltonian (1), the second term describes
all the leads, and $H^{LS}$ couples the leads to the plaquette with the 
strength $\tau$. 
With $G^+_{\alpha,\beta}(E)\equiv<\alpha|(E-H+i0)^{-1}|\beta>$, 
the electron transmittance 
 between the leads $\alpha$ and $\beta$, 
in the Landauer-B\"{u}ttiker formalism, is given by: 
\begin{equation}
T_{\alpha,\beta}=4\tau^4 |G^+_{\alpha,\beta}(E_F)|^2 Im g^L_{\alpha}(E_F) 
Im g^L_{\beta}(E_F),
\end{equation}
where $g^L$ is the Green function of the isolated leads.
\begin{figure}[htm]
	\includegraphics[angle=-0,width=0.5\textwidth]{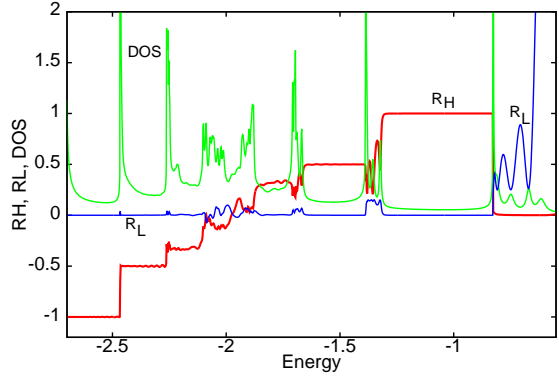}
\caption{(Color online)
The transport properties of the Lieb lattice under the magnetic field: 
Hall resistance $R_H$, longitudinal resistance $R_L$ and density of 
states DOS for a finite Lieb lattice connected to four
transport leads. The quantum Hall effect can be observed for 
$E\in[-2.75,-0.8]$ ($R_H$ integer and $R_L=0$).
In the Bloch-Landau part of the spectrum ($E<-2$)
one has $R_H<0$, while in the Dirac-Landau part ($E>-2$) one has $R_H>0$.
In the energy range $E\in [-0.8,-0.6]$ the transport properties are due 
to the twisted edge states and we get zero Hall resistance $R_H=0$ and 
oscillations of the longitudinal resistance with the characteristic 
minima at $R_L=1/4$. The density of states exhibits maxima at the
transition between the Hall plateaus and for the energy values where the 
twisted edge states appear. The dimension of the plaquette is
$10\times 30$ unit cells, the magnetic flux is $\phi=0.12$.
The resistance is in units $h/e^2$, DOS  in arbitrary units, 
and the energy in units $t$.}
\end{figure}
In what follows we shall discuss the
interesting question of the contribution to  transport of the
twisted edge states introduced in the previous section.
The answer can be found in Fig.15 in the energy range $E\in[-0.8,-0.6]$, 
where one observes that  the twisted edge states found in that range 
do not support the Hall resistance ($R_H=0$), however they contribute to the 
longitudinal resistance, which exhibits an oscillating behavior. 
It is also to  notice  that all the oscillations  minima 
equals $R_L=1/4$, a fact that should find its explanation.

In exploring these unexpected effects, the first step should be to 
identify the transmittance matrix. The numerical investigation  
presented in Fig.17a shows that, in the range of the twisted edge states 
the properties of  transmittances  $T_{\alpha,\beta}$
are very specific: they are not quantized, show an oscillating dependence 
on the energy, and, mainly, satisfy the symmetry relation:
\begin{equation}
T_{\alpha,\alpha+1}=T_{\alpha+1,\alpha}, 
\end{equation} 
while in the range of the conventional edge states, where the quantized  
plateaus occur, the usual properties of quantum Hall effect hold:
$T_{\alpha,\alpha+1}$=integer and  $T_{\alpha+1,\alpha}=0$ 
(for a given sign of the magnetic flux). 
\begin{figure}[htm]
	\includegraphics[angle=-00,width=0.4\textwidth]{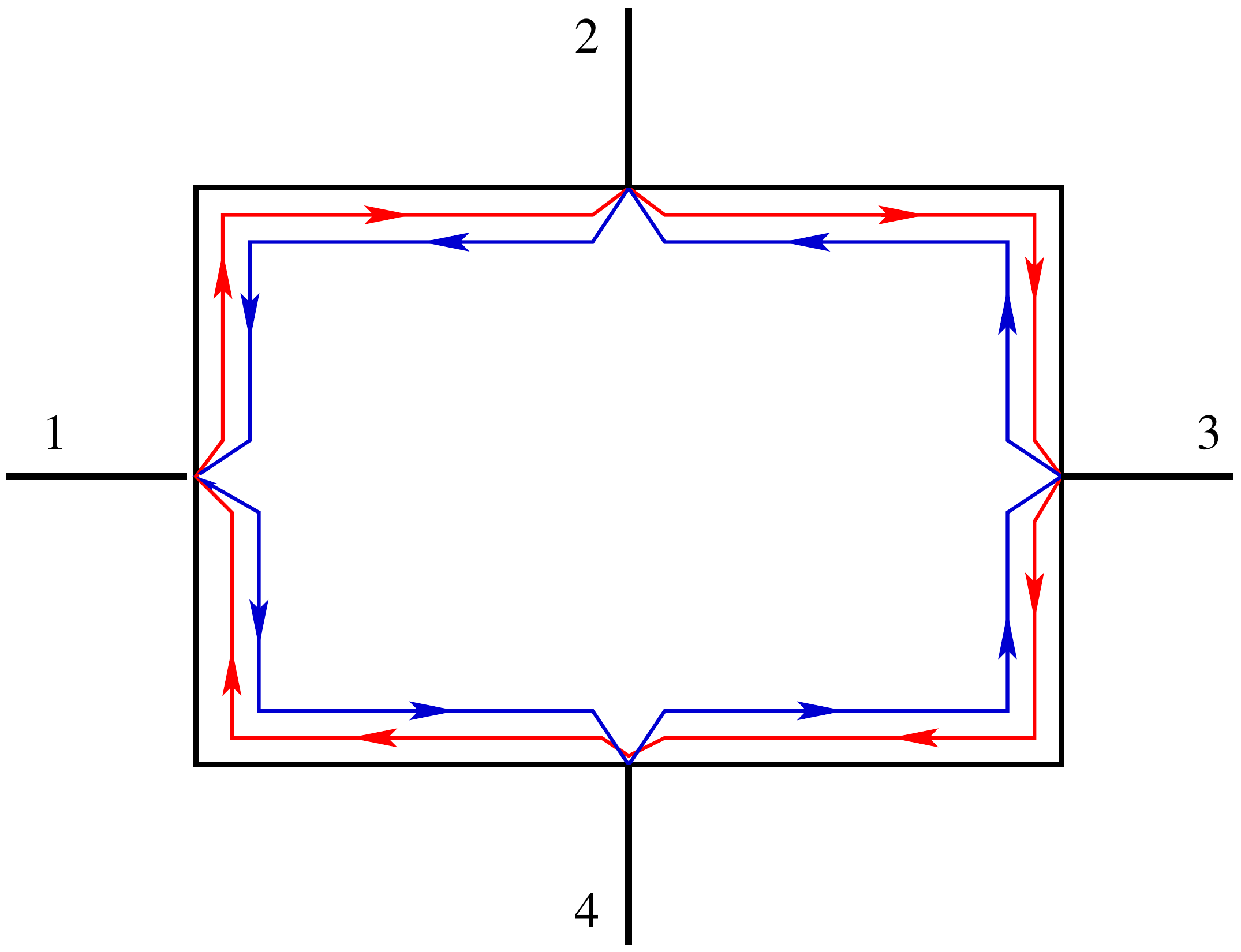}
\caption{(Color online) The four-lead  Hall device: illustration of the 
edge currents carried  by the twisted edge states, for which the 
symmetry relation $T_{\alpha,\alpha+1}=T_{\alpha+1,\alpha}$  holds.}
\end{figure}
Combining Eq.(24)  with the general property 
$\sum_{\alpha} T_{\alpha,\beta}=0$,  which expresses the current conservation, 
it turns out that the transmittance matrix for the edge transport in the domain of twisted edge states  can be written  as:
\begin{displaymath}
        \mathcal{T}=
        \left(\begin{array}{cccc}
                -2T & T & 0 & T\\
                 T & -2T & T & 0\\
                 0 & T & -2T & T\\
                 T & 0 & T & -2T\\
        \end{array}\right).
\hskip1cm
\end{displaymath}
The transmittance  $T_{\alpha,\beta}$ relates the current through the 
lead $\alpha$ to  potentials at the contact sites $\beta$ as:
\begin{equation}
I_{\alpha}=\frac{e^2}{h}\sum_{\beta}T_{\alpha \beta} V_{\beta},
\end{equation}
and, for the four-lead device considered in Fig.16, 
in the Landauer-B{\"u}ttiker formalism, the Hall and longitudinal resistance 
are given (in units $h/e^2$) by:
\begin{eqnarray}
\nonumber R_L&=&R_{14,23}=(T_{24}T_{31}-T_{21}T_{34})/D, \\
R_H&=&(R_{13,24}-R_{24,13})/2= 
(T_{23}T_{41}-T_{21}T_{43}-T_{32}T_{14}+T_{12}T_{34})/2D, 
\end{eqnarray}
where $D=-4T^3$ is a  subdeterminant of the matrix $\mathcal{T}$.
By the use of the above equations and of transmittance matrix $\mathcal{T}$, 
valid in the range of twisted edge states,
one obtains immediately, a vanishing Hall resistance ($R_H=0$) and 
the longitudinal resistance  $R_L=1/4T$ (in units $h/e^2$).
The minima of $R_L$ observed in Fig.17b correspond to  the maxima of the  
transmittance, and, obviously, the value $R_L=0.25 h/e^2$  
expresses  a perfect conducting one channel transport  with $T=1$.
It turns out that, although carried by edge states, the current shows a 
dissipative character.
\begin{figure}[htm]
	\includegraphics[angle=-0,width=0.4\textwidth]{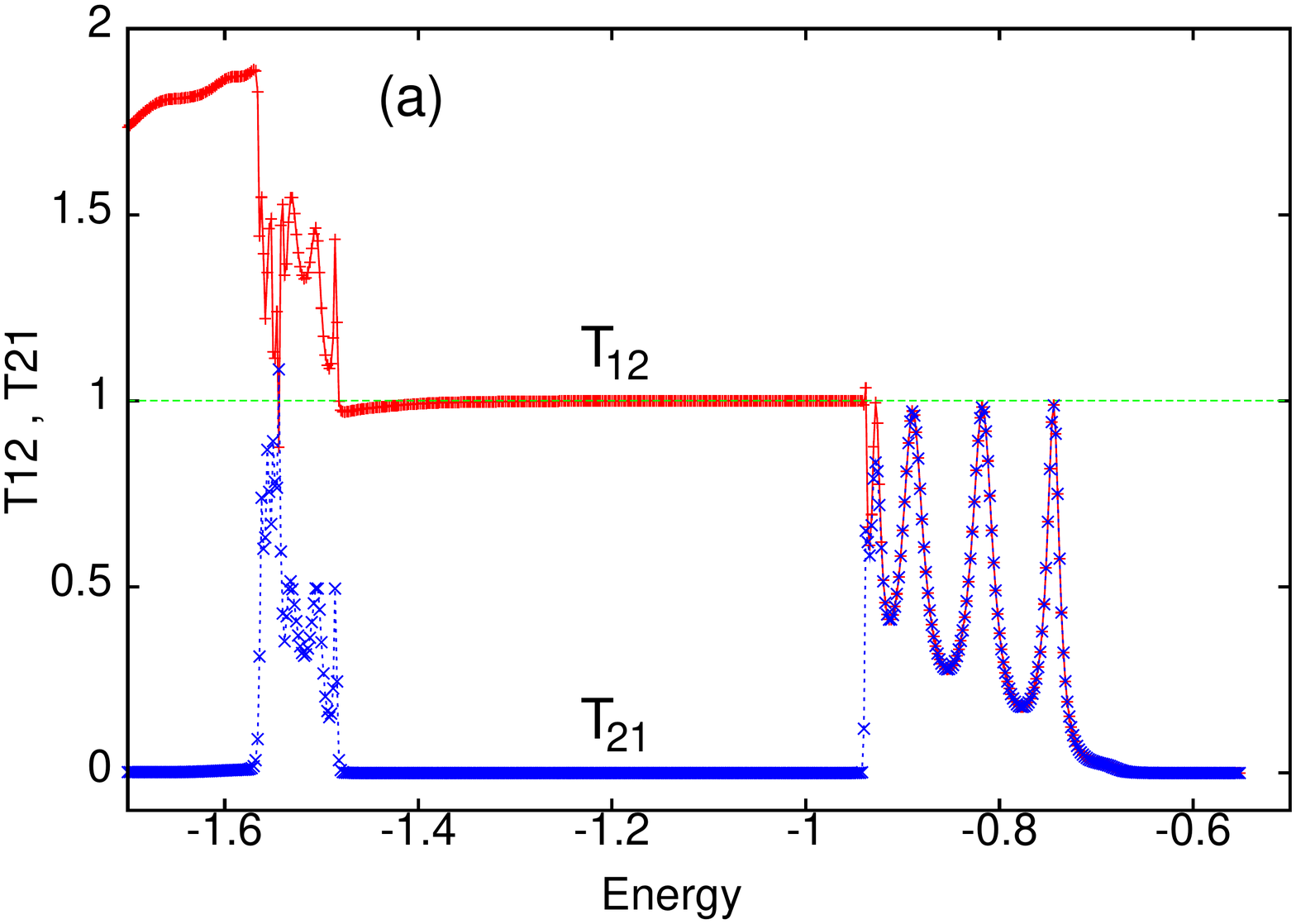}
	\includegraphics[angle=-0,width=0.4\textwidth]{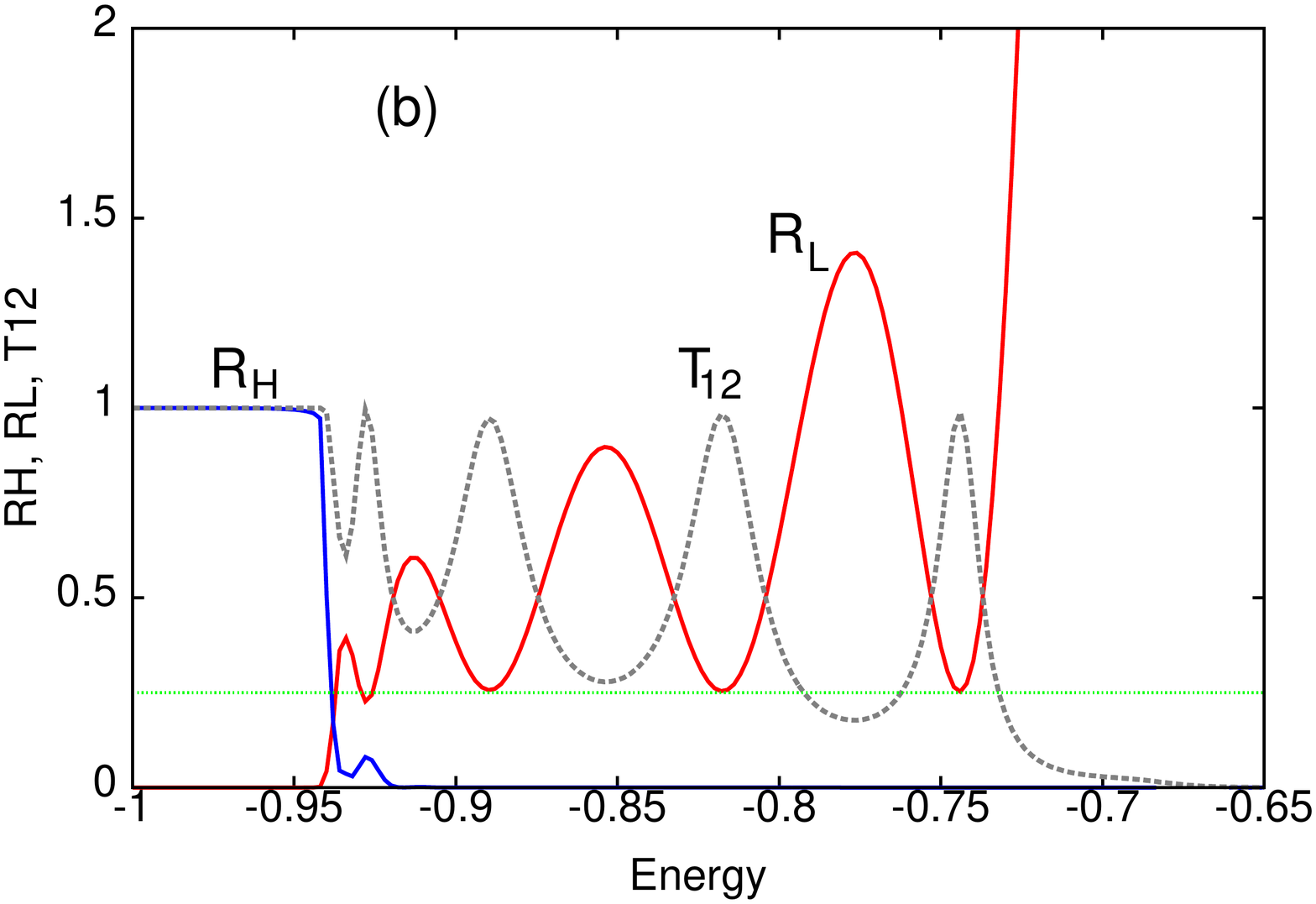}
\caption{(Color online) (a) The transmittances $T_{12}$ and $T_{21}$ 
showing the quantized values $T_{12}=1$ and $T_{21}=0$ in the range of 
the IQHE, non-quantized oscillating values $T_{12}=T_{21}$ in the 
range of twisted edge states $E>-0.95$, and $T_{12}=T_{21}+1$ in the
range of type-II edge states $E\in [-1.55,-1.5]$.
(b) The Hall and longitudinal resistance:   
$R_H$ (blue line) vanishes   in the range of twisted edge states,
while $R_L$ (red line) exhibits oscillations with minima $R_L=0.25$. 
The  minima of the longitudinal resistance occur at the energies where the 
transmittance (black dashed line) get the maximum value $T_{12}=T_{21}=1$.
The dimension of the plaquette is $10\times 30$ unit cells, $\phi=0.16$.} 
\end{figure}
The oscillations of $T_{12}$ and $T_{21}$ in the domain of the 
twisted edge states ($E \in [-0.95,-0.75]$) follow the similar 
oscillations of the density of states,  while in the quantum Hall 
regime (E $\in$ [-1.5,-0.95])  the DOS is flat. 

Another interesting problem is the transition between the first and second
plateau of the Dirac-Landau region ($E \in [-1.55,-1.5]$) which is
much  wider than  similar transitions in the Bloch-Landau region. 
The transition get a width which is due to the presence of the type-II
edge states (observed in the spectrum Fig.12 above the Dirac band),
and is accompanied by oscillations of the transmittance 
(see $T_{12}$ and $T_{21}$ in Fig.17a). We have to stress that
in this energy range $T_{12}$ and $T_{21}$ are no more equal, and the
numerical calculation suggests that $T_{12}=T_{21}+1$. It means that the
symmetry Eq.(24) remains specific to the twisted edge states.

In order to figure out a scenario for the vanishing Hall effect,
we remind  first that, in the range of the  QH effect, 
all the (conventional) edge states responsible for the plateaus of the
transverse magnetoresistance get a unique chirality determined by the 
direction of the magnetic flux. This results in a definite sense 
(say clock-wise) of the current such that $T_{\alpha,\alpha+1}$= integer 
and $T_{\alpha+1,\alpha}=0$.
The symmetry $T_{\alpha,\alpha+1}=T_{\alpha+1,\alpha}$ which occurs in the 
range of the twisted edge states is characteristic to the absence
of the magnetic field. So, this symmetry indicates a 'loss of influence' of 
the magnetic field followed by a vanishing Hall effect. 
As a support of this idea we  note that  the twisted edge states show 
alternating (clock and anti-clock) chiralities which allow for the 
transmittance in both directions, as sketched in Fig 16. 
This might be  an heuristic explanation for the vanishing of the 
transverse resistance despite the presence of edge states. 
\begin{figure}[htm]
\includegraphics[angle=-0,width=0.45\textwidth]{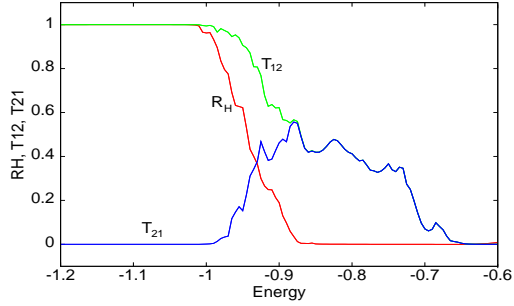}
\caption{(Color online)
The Hall resistance $R_H$ (red) and transmittances $T_{12}$ (green) and 
$T_{21}$ (blue) for a disordered Lieb lattice  in the 
 transition region from the first Hall plateau to  the domain of
 twisted edge states.
The figure shows that the symmetry relation $T_{12}=T_{21}$ for 
twisted edge state transport holds also in the  presence of disorder.
The Hall resistance is $R_H=1$ for the energies corresponding to the 
first gap with conventional edge states, 
and $R_H=0$ for the energies corresponding to the twisted edge states.
The dimension of the plaquette and the magnetic flux are the same as in Fig.17,
the disorder amplitude is $W=0.5$.}
\end{figure}

Let us discuss now the effect of the disorder. It is known that
the conventional edge states are robust to disorder, whereas the bulk states
(which form the Landau bands) are more sensitive, so that the IQHE of a 
disordered plaquette shows robust plateaus, and a broadened transition 
region between two consecutive plateaus.
On the other hand, as we have shown, the disorder localizes easily the
twisted edge states, changing in this way their transmittance properties.
Fig.18  depicts the transmittances $T_{12},T_{21}$ for a disordered Lieb 
plaquette compared to the same transmittances of the clean system. 
One observes  the  quantized values in the range of the 
conventional edge states followed, in the range of twisted states,
by  reduced values of the disordered transmittance which replace the 
peaks specific to the clean system. It is worth to say  that the symmetry 
$T_{\alpha,\alpha+1}=T_{\alpha+1,\alpha}$ is preserved 
for each individual disordered sample, and, as a consequence,
the Hall effect vanishes similar to the  clean case. 

\section{Conclusions}
We have found that the specific topology of the  2D Lieb lattice induces 
remarkable spectral and transport properties. Up to point there are 
similarities with the electronic energy spectrum of graphene in what concerns 
the presence of a Dirac-type cone at low energy, however, in addition, 
a macroscopically degenerated  flat band occurs at the middle of the spectrum. 
The perpendicular magnetic field applied on a finite (mesoscopic) Lieb 
plaquette opens a gap around the flat band, and we show the presence 
in this gap of a new class of edge states with alternating chirality 
(which we call twisted edge states).
The flat band is insensitive to the magnetic field, however 
an in-plane electric field, and also the disorder,  lifts the degeneracy. 
The electric field applied on a finite (mesoscopic) system gives rise to a
Wannier-Stark fan composed of degenerate mini-bands, 
the number of them being equal to the number of cells along  
the direction of the field. 

The macroscopic degeneracy of the flat band is lifted by  disorder,
and the degree of localization and the level spacing distribution are
studied.  It turns out that not only the ordered  flat band, but also the
disordered one, does not feel the magnetic field; indeed, we prove that
the level spacing distribution of the disordered system follows the
orthogonal ($\beta=1$) Wigner-Dyson distribution, which usually describes 
disordered systems in the absence of the magnetic field.

We calculate analytically the orthogonal eigenfunctions of the 
finite Lieb system corresponding to the three spectral branches 
in the low energy range, both for periodic and vanishing  boundary conditions.
In this way we find also the degeneracy of the zero energy flat band which,
in the periodic case, equals the total number of unit cells $N_{cell}$ 
(except  when both $N^x_{cell}$ and $N^y_{cell}$ are even numbers,
in which case the degeneracy increases to  $N_{cell}+2$), while in the
case of the closed boundaries the degeneracy is $N_{cell}+1$.
A toy model composed of only two unit cells helps to understand the
behavior in the presence of a perpendicular magnetic field. The perturbative
calculation shows that 2 states of the flat band separate from the
degenerated bunch, and belong to the  class of twisted edge states. 

The eigenenergies of the twisted edge states depend in an oscillatory 
manner on the magnetic flux, i.e., show an alternating chirality, 
and contrary to the conventional edge states, the diamagnetic moment 
change the sign when the magnetic flux is varied. 
These type of edge states generated by the magnetic field are not protected 
by the broken time-reversal symmetry, and proves to get localized even at 
low disorder, when the conventional edge states remain robust.

The transport properties  are calculated by attaching
leads to the finite Lieb system and using the Landauer-B{\"u}ttiker formalism.
The quantum  Hall resistance looks similar to that of the graphene except 
the steps are equal to $h/e^2$ (instead of $h/2e^2$), however  in the domain
of the twisted edge states the properties become unconventional:
the Hall resistance vanishes, while the longitudinal one shows oscillations
which can be correlated with the oscillations of the density of states 
(calculated in the presence of the leads). This behavior stems from the 
symmetry of the transmittance $T_{\alpha,\alpha+1}=T_{\alpha+1,\alpha}$ which
occurs despite the presence of the quantizing magnetic field. The symmetry holds also in the presence of disorder.

\section{Acknowledgements}
We acknowledge  support from PNII-ID-PCE Research Programme 
(grant no 0091/2011), Core Programme (contract no.45N/2009) and
Sonderforschungsbereich 608 at the  Institute of Theoretical Physics,
University of Cologne. One of the authors (AA) is very much indebted 
to A.Rosch for illuminating discussions.


\begin{references}
\bibitem{note3} This is to be expected since also the corresponding Landau 
	bands show opposite chirality.
\bibitem{note1} The energy $E_a=E_b=E_c=E_0$ in the Hamiltonian (1) 
is a random variable described by the distribution $P(E)=1/W$ 
for $-W/2<E<W/2$, and zero otherwise.
\bibitem{note2} The degeneracy  at the apparent 
crossing points could be shown only numerically; by reducing 
progressively the grid down to $\Delta\Phi=0.00002$ (in units $\phi_0$)
the distance  between the consecutive eigenvalues decreases 
progressively to $\Delta E=0.00004$ (in units $t$).
\end{references}
\end{document}